\documentclass[a4paper,11pt]{article}
\pdfoutput=1
\usepackage{jheppub}
\makeatletter
\gdef\@fpheader{\relax}
\makeatother
\renewcommand\afterLogoSpace{\vskip-\baselineskip}
\renewcommand\afterSubheaderSpace{}
\renewcommand\afterProceedingsSpace{}
\renewcommand\afterTitleSpace{\vskip8pt}
\renewcommand\afterRuleSpace{\vskip18pt}
\renewcommand\afterEmailSpace{\vskip22pt}
\usepackage{amsmath,amssymb}
\usepackage{bbm}
\usepackage{booktabs}
\usepackage{graphicx}

\newcommand{\fq}{\mathsf{Q}}
\newcommand{\cC}{\mathcal{C}}
\newcommand{\cH}{\mathcal{H}}
\newcommand{\cO}{\mathcal{O}}
\newcommand{\cR}{\mathcal{R}}
\newcommand{\tr}{\operatorname{Tr}}

\title{Twist and Shout: defect lines on the R\'enyi torus}

\author[a]{Naveen Balaji Umasankar}
\affiliation[a]{Department of Physics, Yale University, 217 Prospect St, New Haven, CT 06511, USA}
\emailAdd{naveen.umasankar@yale.edu}

\abstract{
We consider what the entanglement between two regions of a critical system can reveal about its symmetries, including non-invertible ones. For two intervals in a two-dimensional conformal field theory, the second R\'enyi entropy is a partition function on a torus, and a topological defect line can be wrapped around one of the intervals. For an on-site symmetry this computes the overlap of the reduced density matrix with its image under the symmetry acting on one interval, and in general it defines an observable that depends only on the cross-ratio. As the intervals approach each other, the observable vanishes at a rate set by the lightest operator on which the line can end, and its monodromy is fixed by the spins of such operators. Since these spins are constrained by the fusion category, the observable detects 't~Hooft anomalies and distinguishes categories with the same fusion rules. It can be computed in closed form for a $\mathbb{Z}_2$, a $\mathbb{Z}_3$ and a Fibonacci line of the $(E_8)_1$ theory, and for the spin-flip and Kramers--Wannier lines of the Ising model, where the spin-flip result agrees with an exact computation in the critical Ising chain. Averaged over a symmetry, the observable reduces to a known Jones-index bound on the mutual information. We also show that the associated twist-field crossing equations are a form of the modular bootstrap, and discuss whether the full fusion category can be recovered from entanglement on surfaces of higher genus.
}

\begin{document}
\maketitle
\flushbottom

\section{Introduction}\label{sec:intro}

The entanglement entropy of an interval of length $\ell$ in the ground state of a two-dimensional conformal field theory is $\frac c3\log(\ell/\epsilon)$, where $c$ is the central charge and $\epsilon$ a short-distance cutoff \cite{Calabrese:2004eu}, and the R\'enyi entropies have the same form. A single interval therefore carries no information about the theory beyond its central charge. The reason is that two points have no conformal invariant, and correspondingly the replica surface of one interval is a sphere. Two disjoint intervals $A=[u_1,v_1]$ and $B=[u_2,v_2]$ are the simplest configuration with an invariant, the cross-ratio
\[
x=\frac{(v_1-u_1)(v_2-u_2)}{(u_2-u_1)(v_2-v_1)}\in(0,1),
\]
which is small when the intervals are far apart and tends to one when they become adjacent. Their mutual information is a finite function of $x$ that depends on the operator dimensions and OPE coefficients of the theory \cite{Furukawa:2008uk,Calabrese:2009ez,Calabrese:2010he,Headrick:2010zt}.

Let $\rho_{AB}$ be the reduced density matrix of $A\cup B$ in the ground state. For integer $n$, $\tr\rho_{AB}^n$ is the partition function on $n$ copies of the plane glued cyclically along the intervals. For $n=2$ this surface is a torus, and $\tr\rho_{AB}^2$ is, up to a universal factor, the torus partition function of the theory at a modulus determined by $x$ \cite{Lunin:2000yv,Headrick:2010zt}. We restrict to $n=2$ throughout, and we make no claims about the von Neumann entropy.

Our aim is to combine this fact with the description of symmetries in terms of topological defect lines \cite{Frohlich:2006ch,Bhardwaj:2017xup,Chang:2018iay}. A topological line is a line operator that commutes with the stress tensor, so that it can be deformed freely as long as it does not cross other operators. The elements of a symmetry group correspond to invertible lines. In general a two-dimensional theory also has lines that are not invertible, such as the Kramers--Wannier duality line of the Ising model, and the set of lines is described by a fusion category. On a torus a line can be wrapped on either of the two cycles, and we would like to understand what the resulting partition functions mean when the torus is the replica surface of two intervals.

The interpretation is simple. One of the cycles of the replica torus is a loop that encircles $A$ on one of the two sheets. A line $L$ on this loop acts on one of the two factors of $\rho_{AB}$ in $\tr\rho_{AB}^2$, and if $L$ is the line of a group element $g$ that acts on-site it conjugates that factor by $g_A$, the restriction of $g$ to $A$. The ratio of the partition functions with and without the line is then the normalized overlap between $\rho_{AB}$ and $g_A\rho_{AB}g_A^{-1}$. For a general line, which may be non-invertible or generate an anomalous symmetry, we take this ratio of partition functions as the definition of an observable $\cO_L(x)$. It measures the charge exchanged between $A$ and $B$, since it would be trivial if $\rho_{AB}$ commuted with the symmetry acting on $A$ alone. No prescription is needed for the symmetry operator at the endpoints of $A$, because the loop is closed, and consequently $\cO_L$ depends only on $x$. Under the modular transformation $S$, which exchanges $x$ and $1-x$, the line is mapped to the other cycle, where it changes the Hilbert space on which the trace is taken. The behaviour of $\cO_L$ for nearly adjacent intervals is therefore governed by the defect Hilbert space of $L$, that is, by the operators on which $L$ can end.

Some of this is known. If $\cO_L$ is averaged over the lines of a symmetry with suitable weights, one obtains the difference between the mutual information of $A$ and $B$ in the given theory and in its subtheory of neutral operators. For a finite group $G$ this difference was studied in \cite{Casini:2019kex}, where it was shown to be a relative entropy, to be bounded by $\log|G|$, and to saturate the bound when the two regions touch. The extension to rational conformal field theories, in which $\log|G|$ is replaced by a quantity determined by a Jones index, was given in \cite{Benedetti:2024dku}, and the Ising model was treated at finite R\'enyi index in \cite{Benedetti:2026jones}. The averaged quantity defined in section~\ref{sec:avg} is the $n=2$ case of the one considered in these papers. There is also previous work on two intervals with symmetry operators inserted in the replica construction. For U(1) symmetries these include the multi-charged moments of \cite{Ares:2022gjb}, in which magnetic fluxes end at the branch points, and the charged moments of the realignment matrix \cite{Bruno:2023tez}, which are given by torus partition functions with a charged loop on a non-contractible cycle. In addition, in fermionic theories the insertion of fermion parity on the cycles of the replica torus amounts to a choice of spin structure \cite{Coser:2015dvp}. Symmetry operators that end on the entangling points enter the charged moments of symmetry-resolved entanglement \cite{Goldstein:2017bua,Xavier:2018kqb,Saura-Bastida:2024yye,Das:2024qdx} and the entanglement asymmetry \cite{Ares:2022koq}, which has been extended to non-invertible symmetries in \cite{Benini:2025asym}.

The present paper is concerned with $\cO_L$ for an individual line, which may or may not be invertible. We show that for $x\to1$ it vanishes as $(1-x)^{2\Delta_L}$, where $\Delta_L$ is the lowest scaling dimension in the defect Hilbert space $\cH_L$, and that its monodromy around $x=1$ is determined by the spins of the states in $\cH_L$. These spins obey selection rules that depend on the associativity data of the category and not only on its fusion rules \cite{Chang:2018iay,Lin:2019kpn}, and through the monodromy they become visible in the replica observable. It detects the 't~Hooft anomaly of a $\mathbb{Z}_N$ symmetry, and it distinguishes the Kramers--Wannier line of the Ising model from the line of $SU(2)_2$ with the same fusion rules.

We then evaluate $\cO_L$ in two theories. The first is the $(E_8)_1$ WZW model, whose chiral algebra has a single representation and whose lines can be constructed from pairs of commuting subalgebras \cite{Umasankar:2026tdl}. For a $\mathbb{Z}_2$ and a $\mathbb{Z}_3$ line the overlap is a rational function of a Hauptmodul of $\Gamma_0(2)$ and of $\Gamma_0(3)$ respectively, and for the Fibonacci line associated with the subalgebra $(G_2)_1\times(F_4)_1$ it is given by hypergeometric functions of a variable on which $S$ acts by a sign. This theory has also been proposed to have lines forming the Haagerup fusion category \cite{Albert:2025haag}, to which we return in section~\ref{sec:disc}. The second theory is the Ising model, where $\cO_L$ is an algebraic function of $x$ for the spin-flip line and for the Kramers--Wannier line, and where we confirm the spin-flip result by an exact computation in the critical transverse-field chain. Finally, $\tr\rho_{AB}^2$ is also a four-point function of twist fields in the $\mathbb{Z}_2$ orbifold of two copies of the theory. We expand this correlator, with and without lines, in Virasoro blocks of central charge $2c$, using the recursion relation for the blocks as in \cite{Ruggiero:2018hyl} and the software of \cite{Poland:2026vb}, and we show that the resulting crossing equations are a form of the modular bootstrap with symmetry lines \cite{Lin:2019kpn,Lin:2021udi}.

The paper is organized as follows. In section~\ref{sec:replica} we review the replica torus and topological lines. In section~\ref{sec:loop} we define $\cO_L$ and its average over a fusion category and derive their general properties. Sections~\ref{sec:E8} and~\ref{sec:ising} contain the results for $(E_8)_1$ and for the Ising model, including the lattice computation. In section~\ref{sec:anom} we show what the monodromy at $x=1$ reveals about the category, with $SU(2)_1$ and free fermions as examples. Section~\ref{sec:blocks} deals with the expansion in Virasoro blocks, and section~\ref{sec:disc} contains a summary and some open questions. The appendices contain a derivation of \eqref{eq:LM}, formulas for characters, tables of block coefficients and details of the lattice computation.

\section{The replica torus and topological lines}\label{sec:replica}

\subsection{The R\'enyi-2 entropy of two intervals}

\begin{figure}[!htb]
\centering
\includegraphics[width=\textwidth]{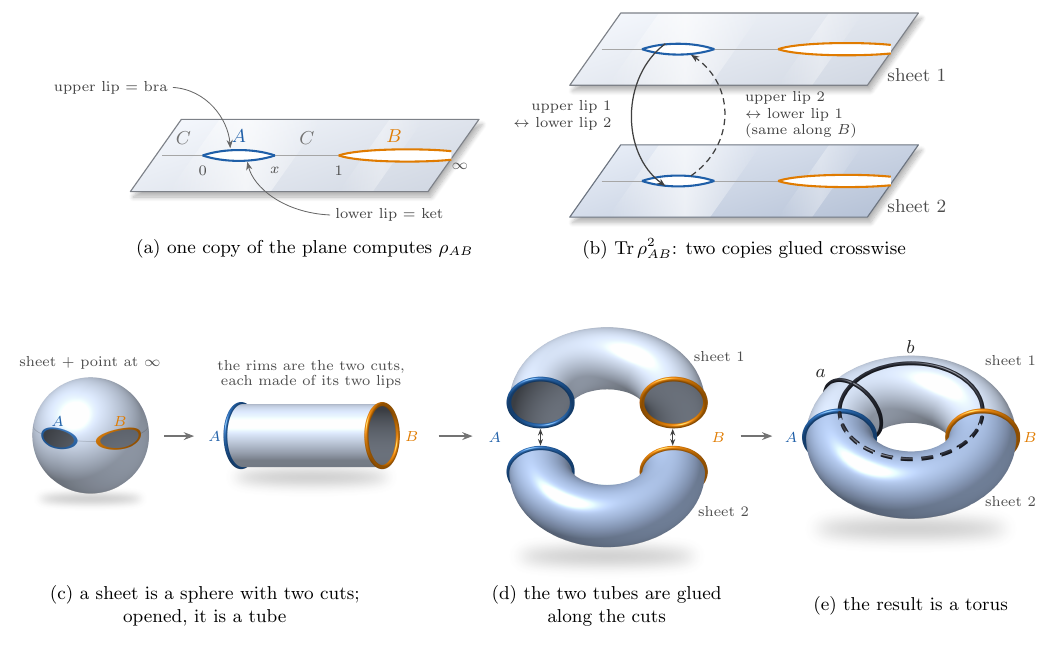}
\caption{The replica surface for $\tr\rho_{AB}^2$. (a) The path integral on the plane cut along $A$ and $B$ gives the matrix elements of $\rho_{AB}$, with the two sides of each cut carrying the two indices. (b) Two such planes glued along the cuts. (c) With the point at infinity added, a sheet is a sphere with two cuts, and when the cuts are opened it becomes a tube whose ends are the cuts $A$ and $B$. (d) The gluing joins the two tubes along these ends. (e) The resulting surface is a torus, on which the cuts are two circles separating the sheets. The $a$-cycle goes around the tube on one sheet, parallel to the cut $A$. The $b$-cycle runs from $A$ to $B$ on the first sheet (solid) and back on the second (dashed).}
\label{fig:replica}
\end{figure}

Let $T$ be a unitary conformal field theory with $c=\bar c$\,\footnote{This is assumed for simplicity. For $c\ne\bar c$ the derivation in appendix~\ref{app:replica} gives \eqref{eq:G} with the factor $(2^8x(1-x))^{-c/24}(2^8\bar x(1-\bar x))^{-\bar c/24}$, and the twist fields have weights $c/16$ and $\bar c/16$.} and a unique vacuum, in its ground state on the real line, and let $A=[u_1,v_1]$ and $B=[u_2,v_2]$ be disjoint intervals with complement $C$. The matrix element $\langle\phi_+|\rho_{AB}|\phi_-\rangle$ is given by the Euclidean path integral on the plane cut open along $A$ and $B$, with the fields equal to $\phi_+$ just above the cuts and to $\phi_-$ just below (figure~\ref{fig:replica}a). Taking the trace of the square of this matrix amounts to gluing two such planes, the upper side of each cut on one sheet being identified with the lower side of the same cut on the other (figure~\ref{fig:replica}b). The result is a two-sheeted cover of the sphere branched at the four endpoints.

To see that this surface is a torus, add the point at infinity to each sheet, so that it becomes a sphere, and open the two cuts. A sphere with two holes is a tube. Its two boundary circles are the cuts $A$ and $B$, each circle being made of the two sides of the cut (figure~\ref{fig:replica}c). The gluing joins the two tubes end to end, the circle $A$ of one sheet to the circle $A$ of the other and likewise for $B$ (figure~\ref{fig:replica}d), and two tubes joined at both ends form a torus (figure~\ref{fig:replica}e). On the torus the two cuts are circles around the tube, and they separate the two sheets.

We call the cycle that encircles $A$ on one sheet the $a$-cycle. It goes around the tube and is parallel to the circle $A$. The $b$-cycle encircles the gap between $A$ and $B$. A closed curve around the gap has to cross both cuts, and crossing a cut leads from one sheet to the other, so the $b$-cycle consists of an arc from $A$ to $B$ on one sheet and an arc from $B$ back to $A$ on the other (figures~\ref{fig:replica}e and~\ref{fig:loops}c).

We write the torus partition function of $T$ as
\[
Z(\tau,\bar\tau)=\tr_\cH\fq^{L_0-c/24}\bar\fq^{\bar L_0-c/24},\qquad\fq=e^{2\pi i\tau},
\]
where $\cH$ is the Hilbert space on the circle, $L_0+\bar L_0$ is the scaling dimension $\Delta$ and $L_0-\bar L_0$ the spin. When $\cH$ decomposes into modules $i$ and $\bar\jmath$ of the holomorphic and antiholomorphic chiral algebras, with multiplicities $M_{i\bar\jmath}$, the partition function is a sum of products of characters $\chi_i(\tau)=\tr_i\fq^{L_0-c/24}$,
\[
Z(\tau,\bar\tau)=\sum_{i,\bar\jmath}M_{i\bar\jmath}\,\chi_i(\tau)\,\overline{\chi_{\bar\jmath}(\tau)} .
\] The modulus of the replica torus is determined by the cross-ratio, and \cite{Lunin:2000yv,Headrick:2010zt}
\begin{equation}\label{eq:LM}
\frac{\tr\rho_{AB}^2}{\tr\rho_A^2\,\tr\rho_B^2}=2^{-2c/3}\,x^{c/6}(1-x)^{-c/12}\,Z(\tau,\bar\tau),\qquad x=\frac{\theta_2(\tau)^4}{\theta_3(\tau)^4}.
\end{equation}
Here $\theta_2=\sum_{n\in\mathbb{Z}}q^{(n+1/2)^2}$, $\theta_3=\sum_nq^{n^2}$ and $\theta_4=\sum_n(-1)^nq^{n^2}$ are the Jacobi theta functions of the nome $q=e^{i\pi\tau}$. We will use both $q$ and $\fq=q^2$. Explicitly $\tau=i\mathbf K(1-x)/\mathbf K(x)$, where $\mathbf K$ is the complete elliptic integral of the first kind, and $q=x/16+O(x^2)$. Equation \eqref{eq:LM} is derived in \cite{Lunin:2000yv,Headrick:2010zt}, and in appendix~\ref{app:replica} we give a short derivation based on the conformal Ward identity, which also covers the insertion of a topological line. The logarithm of the left-hand side of \eqref{eq:LM} is the R\'enyi-2 mutual information $I_2(x)=S_2(A)+S_2(B)-S_2(A\cup B)$, in which the cutoff dependence of the individual terms cancels.

For $0<x<1$ the modulus $\tau$ is purely imaginary. As $x\to0$ it tends to $i\infty$, the $a$-cycle is short compared to the $b$-cycle, and $Z$ is dominated by the states of lowest energy on the $a$-cycle. As $x\to1$ the roles of the two cycles are exchanged. The two regimes are related by the modular transformation $S$, which acts as $\tau\to-1/\tau$ and $x\to1-x$.

Equivalently, $\tr\rho_{AB}^2$ is a four-point function in the orbifold $(T\otimes T)/\mathbb{Z}_2$ of twist fields $\sigma$ inserted at the endpoints, around which the two copies are exchanged. The twist field has weights $h_\sigma=\bar h_\sigma=c/16$, and if $A$ and $B$ are mapped to $[0,x]$ and $[1,\infty)$,
\begin{equation}\label{eq:G}
G(x,\bar x)\equiv\langle\sigma(\infty)\sigma(1)\sigma(x,\bar x)\sigma(0)\rangle=\left|2^8x(1-x)\right|^{-c/12}Z(\tau,\bar\tau).
\end{equation}
This description will be used in section~\ref{sec:blocks}.

\subsection{Topological defect lines}

A topological defect line $L$ is a line operator that commutes with both components of the stress tensor. Placed on a spatial circle it defines an operator $\hat L$ on $\cH$ that commutes with the two Virasoro algebras. On the vacuum $\hat L$ has eigenvalue $d_L$, the quantum dimension of $L$. In a unitary theory $d_L\ge1$, with equality if and only if $L$ is invertible. The invertible lines form the symmetry group $G$ of the theory, with $\hat L_g\hat L_h=\hat L_{gh}$. In general the simple lines obey $\hat L_a\hat L_b=\sum_cN_{ab}^{\ c}\hat L_c$ with non-negative integers $N_{ab}^{\ c}$. An example that we will meet below is the Fibonacci line $W$, with $W\times W=1+W$ and $d_W=\varphi=(1+\sqrt5)/2$. Reversing the orientation of $L$ gives a line $\bar L$, which acts as $\hat L^\dagger$ and has $d_{\bar L}=d_L$. We consider a finite set of simple lines closed under fusion and orientation reversal. They are the simple objects of a fusion category $\cC$, whose total dimension is
\[
\dim\cC=\sum_a d_a^2 .
\]
For the category of a finite group this is the order of the group.

\begin{figure}[!htb]
\centering
\includegraphics[width=\textwidth]{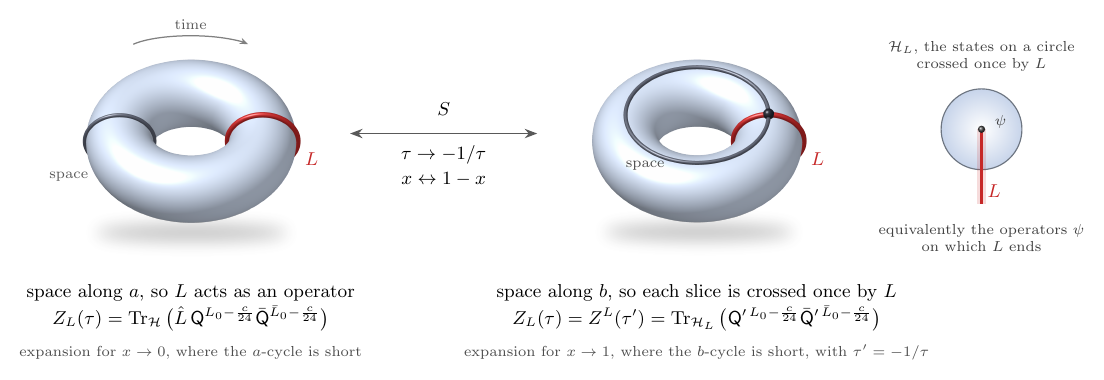}
\caption{A line $L$ on the $a$-cycle of a torus (red), quantized in two ways, with a spatial slice shown in gray. If the $a$-cycle is taken as space, $L$ acts as an operator on $\cH$ and the partition function is $Z_L$. If the $b$-cycle is taken as space, each spatial slice is crossed once by $L$, and the trace is over the defect Hilbert space $\cH_L$. The two descriptions are related by $S$, under which $x\to1-x$.}
\label{fig:torus}
\end{figure}

The torus partition function with $L$ along the spatial cycle is
\[
Z_L(\tau,\bar\tau)=\tr_\cH\big(\hat L\,\fq^{L_0-c/24}\bar\fq^{\bar L_0-c/24}\big).
\]
If $L$ is instead along the time direction, it intersects each spatial slice in one point (figure~\ref{fig:torus}), and the states on such a slice form the defect Hilbert space $\cH_L$. By the state-operator correspondence $\cH_L$ is the space of operators at which $L$ can end. Its partition function $Z^L=\tr_{\cH_L}\fq^{L_0-c/24}\bar\fq^{\bar L_0-c/24}$ is a sum of products of characters with non-negative integer coefficients, and it is related to the previous one by $Z_L(\tau)=Z^L(-1/\tau)$ \cite{Chang:2018iay}. The spins of the states in $\cH_L$ need not be integers. Their values modulo one are constrained by $\cC$ \cite{Chang:2018iay}.

\section{A symmetry acting on one interval}\label{sec:loop}

\subsection{The observable}

\begin{figure}[!htb]
\centering
\includegraphics[width=\textwidth]{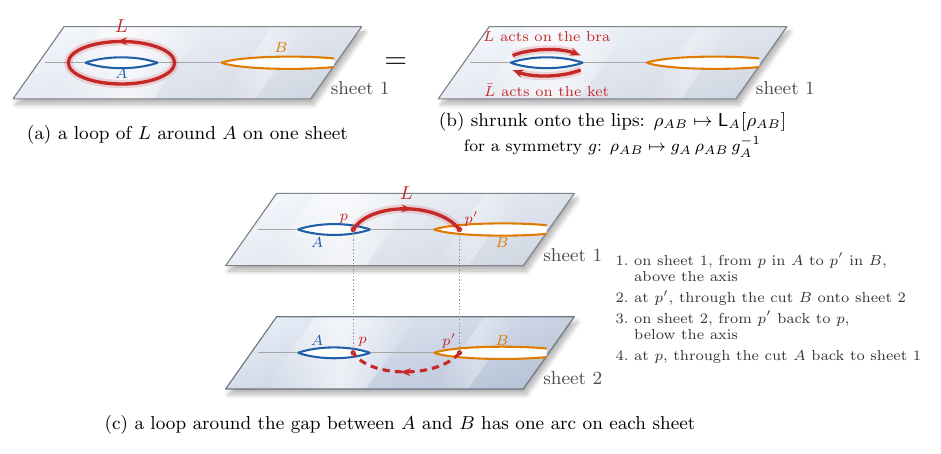}
\caption{Lines on the replica surface. (a) A loop of $L$ around $A$ on the first sheet, which is the $a$-cycle of the torus. (b) The loop contracted onto the cut, where $L$ acts on one index of the first factor $\rho_{AB}$ and $\bar L$ on the other. For an on-site symmetry this is conjugation by $g_A$. (c) The $b$-cycle, a loop around the gap between the intervals. On the first sheet the line runs above the axis from a point $p$ of $A$ to a point $p'$ of $B$. There it passes through the cut onto the second sheet, on which it returns below the axis from $p'$ to $p$, and it comes back to the first sheet through the cut $A$. Seen from above, the two arcs form a closed curve that encircles the gap.}
\label{fig:loops}
\end{figure}

Consider now a line $L$ along the $a$-cycle of the replica torus, that is, on a loop that encircles $A$ on the first sheet (figure~\ref{fig:loops}a). We define $\cO_L$ as the ratio of the partition functions on the replica surface with and without this loop. The factor that relates the replica partition function to $Z$ in \eqref{eq:LM} is not affected by the insertion of a topological line, and therefore
\begin{equation}\label{eq:obs}
\cO_L(x)\equiv\frac{Z_L(\tau,\bar\tau)}{Z(\tau,\bar\tau)},\qquad \cR_L(x)\equiv\frac{\cO_L(x)}{d_L}.
\end{equation}
This definition applies to any topological line.

Its meaning in terms of $\rho_{AB}$ is clearest for a symmetry $g$ that acts on-site, as a product of operators on the sites of a lattice regularization. Since $L$ is topological, the loop can be contracted until it runs along the two sides of the cut $A$ (figure~\ref{fig:loops}b). On the first sheet the two sides of the cut carry the two indices of the first factor of $\rho_{AB}$ in $\tr\rho_{AB}^2$, and the loop conjugates this factor by $g_A$, the product of the on-site operators over $A$. Hence
\begin{equation}\label{eq:overlap}
\cO_g(x)=\frac{\tr\big(\rho_{AB}\,g_A\rho_{AB}g_A^{-1}\big)}{\tr\rho_{AB}^2}.
\end{equation}
For other lines the contraction of the loop still replaces the first factor of $\rho_{AB}$ by an object $\mathsf L_A[\rho_{AB}]$, so that $\cO_L=\tr(\rho_{AB}\,\mathsf L_A[\rho_{AB}])/\tr\rho_{AB}^2$, but $\mathsf L_A$ is then not conjugation by an operator on $A$. Non-invertible lines do have lattice realizations. The Kramers--Wannier operator of the Ising chain, for instance, is a well-defined operator on a closed chain \cite{Seiberg:2023cdc}. It is not a product of operators on single sites, however, and it has no canonical restriction to a block, so that its action on $\rho_{AB}$ depends on how the ends of the block are treated. The same is true of an anomalous symmetry, which cannot be made on-site. In these cases we take the loop on the replica torus as the definition. An action of non-invertible symmetries on the density matrix of a subsystem was defined in \cite{Benini:2025asym}, where it depends on a choice of boundary conditions at the endpoints of the subsystem. The loop considered here does not pass through the endpoints, and it would be interesting to understand how the two constructions are related.

In \eqref{eq:overlap} the ground state is invariant under $g$, and hence $\rho_{AB}$ commutes with $g_Ag_B$, but there is no reason for it to commute with $g_A$. It does so if $\rho_{AB}=\rho_A\otimes\rho_B$, and in general $\cO_g$ differs from one when the correlations between $A$ and $B$ involve the exchange of charge. A similar statement holds for any line. If the loop could be contracted to a point it would contribute a factor $d_L$. This is the case when $A$ and $B$ are uncorrelated, since the replica surface of a single interval is a sphere. We have normalized $\cR_L$ so that it is equal to one in this situation.

In the continuum the operator $g_A$ requires a prescription at the endpoints of $A$, and quantities that involve it generally depend on the prescription. This problem does not arise for $\cO_L$. The line in figure~\ref{fig:loops}a is closed and does not pass through the branch points, and the cutoff-dependent normalization of $\tr\rho_{AB}^2$ cancels in the ratio \eqref{eq:obs}. Thus $\cO_L$ is a function of the cross-ratio only. For a U(1) symmetry, \eqref{eq:overlap} inserts $e^{i\alpha Q_A}$ after the first factor of $\rho_{AB}$ and $e^{-i\alpha Q_A}$ after the second, where $Q_A$ is the charge in $A$. The corresponding magnetic fluxes cancel around each branch point, and what remains is a loop on a non-contractible cycle of the replica torus. This distinguishes $\cO_g$ from the multi-charged moments of two intervals \cite{Ares:2022gjb}, which contain a single insertion of the charge operators and in which the fluxes end at the branch points, where they are represented by vertex operators. Torus partition functions with a charged loop on a non-contractible cycle appear instead in the symmetry resolution of the realignment negativity \cite{Bruno:2023tez}.

\subsection{Small and large cross-ratio}\label{sec:limits}

The behaviour of $\cO_L$ at small and at large cross-ratio follows from the two descriptions of figure~\ref{fig:torus}, each of which leads to a low-temperature expansion. With the $a$-cycle as space, the length of the $b$-cycle plays the role of an inverse temperature, and it becomes large as $x\to0$. With the $b$-cycle as space, the roles are exchanged, and the inverse temperature becomes large as $x\to1$. For $x\to0$ we have $\fq\simeq(x/16)^2\to0$, and \eqref{eq:obs} can be expanded in the states of $\cH$. The vacuum contributes $d_L$ to $Z_L$ and $1$ to $Z$. Let $\psi$ run over the remaining states, with scaling dimensions $\Delta_\psi$ and eigenvalues $\lambda_L(\psi)$ of $\hat L$. To leading order
\begin{equation}\label{eq:small}
\cR_L(x)\simeq1-\sum_\psi\Big(1-\frac{\lambda_L(\psi)}{d_L}\Big)\Big(\frac{x}{16}\Big)^{2\Delta_\psi},
\end{equation}
where only the states of lowest dimension with $\lambda_L(\psi)\ne d_L$ need to be kept. The leading deviation of $\cR_L$ from one at large separation is thus controlled by the lightest operators that are charged under $L$, in the sense that $\hat L$ does not act on them as multiplication by $d_L$.

For $x\to1$ we use the dual modulus $\tau'=-1/\tau$, for which $\fq'\simeq((1-x)/16)^2\to0$, and expand in the states of $\cH_L$ instead. Since $Z_L(\tau)=Z^L(\tau')$ and $Z(\tau)=Z(\tau')$, and since the vacuum dominates $Z(\tau')$,
\begin{equation}\label{eq:large}
\cO_L(x)=\frac{Z^L(\tau')}{Z(\tau')}\simeq n_L\Big(\frac{1-x}{16}\Big)^{2\Delta_L},
\end{equation}
where $\Delta_L$ is the lowest scaling dimension in $\cH_L$ and $n_L$ the number of states with that dimension. If $L$ is simple and different from the identity line, $\cH_L$ contains no state of dimension zero, so $\Delta_L>0$ and $\cO_L$ vanishes as the intervals become adjacent. The exponent is twice the dimension of the lightest operator on which $L$ can end.

The expansion in $\cH_L$ also determines the monodromy of $\cO_L$ around $x=1$, that is, the way in which $\cO_L$ changes when it is continued analytically along a closed path in the complex $x$-plane that encircles $x=1$. Let us decompose the defect partition function according to the spin modulo $\frac12$, writing $Z^L=\sum_\sigma Z^L_\sigma$, where $Z^L_\sigma$ contains the states with $s\in\sigma+\frac12\mathbb{Z}$, and correspondingly $\cO_L=\sum_\sigma\cO_{L,\sigma}$ with $\cO_{L,\sigma}=Z^L_\sigma(\tau')/Z(\tau')$. Now let $x$ go once around $x=1$, with $\bar x$ its complex conjugate. Near $x=1$ the dual nome $q'$ is an analytic function of $1-x$ with a simple zero, so that $q'\to e^{2\pi i}q'$, which is $\tau'\to\tau'+2$. The contribution of a state of spin $s$ to $Z^L$ is multiplied by $e^{4\pi is}$, while $Z(\tau')$ is unchanged, and therefore
\begin{equation}\label{eq:monodromy}
\cO_L=\sum_\sigma\cO_{L,\sigma}\ \longrightarrow\ \sum_\sigma e^{4\pi i\sigma}\,\cO_{L,\sigma}.
\end{equation}
The functions $\cO_{L,\sigma}$ are eigenfunctions of the monodromy, with eigenvalues that depend only on the spins modulo $\frac12$. The overlap itself is in general a sum of eigenfunctions with different eigenvalues, and it is not multiplied by a phase unless all the spins in $\cH_L$ are equal modulo $\frac12$. It is single valued in a neighbourhood of $x=1$ if and only if all states in $\cH_L$ have spin in $\frac12\mathbb{Z}$. Otherwise it has a branch point there, and the eigenvalues of the monodromy are determined by the spins in $\cH_L$. We will encounter both cases in sections~\ref{sec:E8} and~\ref{sec:ising}, and in section~\ref{sec:anom} we return to what the monodromy says about the category.

\subsection{Averaging over the symmetry}\label{sec:avg}

We now consider the average of $\cO_L$ over the lines of a symmetry. Take first a finite group $G$, and let
\[
\rho^{G_A}_{AB}=\frac1{|G|}\sum_gg_A\rho_{AB}g_A^{-1}
\]
be the average of $\rho_{AB}$ over the action of $G$ on $A$. This is the density matrix that reproduces the expectation values of those operators on $A\cup B$ that are invariant under $G$ acting on $A$. Its second R\'enyi entropy exceeds that of $\rho_{AB}$ by
\[
\Delta S_2=S_2(\rho_{AB}^{G_A})-S_2(\rho_{AB})=-\log\frac{1}{|G|}\sum_g\cO_g(x),
\]
which is the R\'enyi-2 form of the entanglement asymmetry of \cite{Ares:2022koq} for the group acting on $A$. The von Neumann analogue of this quantity was studied in \cite{Casini:2019kex}, where it was shown to be the relative entropy between the two states, to be bounded by $\log|G|$, and to approach this bound as the regions become adjacent. We will refer to $\Delta S_2$ as the relative asymmetry.

The generalization to a fusion category $\cC$ is obtained by replacing the group average with the operator
\begin{equation}\label{eq:Pi}
\Pi_\cC=\frac{1}{\dim\cC}\sum_ad_a\hat L_a
\end{equation}
on $\cH$. Using the fusion rules and the identity $\sum_aN_{ab}^{\ c}d_a=d_bd_c$ one finds $\Pi_\cC\hat L_b=d_b\Pi_\cC$, and therefore $\Pi_\cC^2=\Pi_\cC$. Moreover $\Pi_\cC$ is Hermitian, since $\hat L_a^\dagger=\hat L_{\bar a}$ and $d_{\bar a}=d_a$. It follows that $\Pi_\cC$ is the orthogonal projector onto the subspace of $\cH$ on which every $\hat L_a$ acts as multiplication by $d_a$ (see also \cite{Lin:2022dhv}). We call the states in this subspace $\cC$-neutral. For a group they are the invariant states. The argument does not require the fusion rules to be commutative. Note that $\Pi_\cC$ is a projector acting on states. It should be distinguished from the conditional expectation onto the algebra of neutral operators \cite{AliAhmad:2025faces}, which acts on operators, and from the symmetrizer of the density matrix of a subsystem \cite{Benini:2025asym}. For a group all three are given by the average over $G$. We define
\begin{equation}\label{eq:dS}
\Delta S^{\cC}_2(x)\equiv-\log\frac{1}{\dim\cC}\sum_ad_a\,\cO_a(x)=-\log\frac{Z_{\Pi_\cC}(\tau,\bar\tau)}{Z(\tau,\bar\tau)},
\end{equation}
where $Z_{\Pi_\cC}$ is the partition function with $\Pi_\cC$ inserted in the trace. This quantity obeys
\begin{equation}\label{eq:bound}
0\le\Delta S_2^{\cC}(x)\le\log\dim\cC .
\end{equation}
The lower bound holds because $\Pi_\cC$ is a projector that commutes with $L_0$ and $\bar L_0$, so that for real $x$ the sum $Z_{\Pi_\cC}$ contains a subset of the positive terms of $Z$. For the upper bound we write $Z_{\Pi_\cC}(\tau)=(\dim\cC)^{-1}\sum_ad_aZ^a(\tau')$. For real $x$ every term in this sum is positive, and the term with $a=1$ is $Z(\tau')=Z(\tau)$, so that $Z_{\Pi_\cC}\ge Z/\dim\cC$.

The bound \eqref{eq:bound} is not new. In a rational conformal field theory the difference of mutual informations between a theory and a subtheory is governed, in the limit of adjacent intervals, by $\frac12\log\mu$, where $\mu$ is the Jones index of the inclusion and can be written in terms of quantum dimensions \cite{Benedetti:2024dku}. For the subtheory of $\cC$-neutral operators $\frac12\log\mu=\log\dim\cC$. In the Ising model, our $Z_{\Pi_\cC}$ for the $\mathbb{Z}_2$ symmetry and for the full Ising category are the partition functions of the two subtheories considered at finite $n$ in \cite{Benedetti:2026jones}. The description in terms of lines gives in addition the way in which the two ends of \eqref{eq:bound} are approached. From \eqref{eq:small} and \eqref{eq:large},
\begin{equation}\label{eq:dSasym}
\Delta S_2^{\cC}\simeq N_\perp\Big(\frac{x}{16}\Big)^{2\Delta_\perp}\quad(x\to0),\qquad
\Delta S_2^{\cC}\simeq\log\dim\cC-\sum_{a\ne1}d_an_a\Big(\frac{1-x}{16}\Big)^{2\Delta_a}\quad(x\to1),
\end{equation}
where $\Delta_\perp$ is the lowest dimension of a state that is not $\cC$-neutral, $N_\perp$ is the number of such states, and $\Delta_a$, $n_a$ are defined as in \eqref{eq:large} for the line $a$. For a group the limiting value at $x\to1$ is $\log|G|$. For a category with non-invertible lines $\dim\cC$ is in general not an integer. The same number appears in the boundary entropy terms of categorical symmetry resolution \cite{Saura-Bastida:2024yye,Das:2024qdx}.

\subsection{Charged correlations}\label{sec:qi}

The relative asymmetry has two simple interpretations. The first is in terms of the mutual information. For a group, the average over $G$ acting on $A$ does not change $\rho_A$ or $\rho_B$, and therefore
\begin{equation}\label{eq:split}
I_2(A{:}B)=I_2(A{:}B)\big[\rho^{\cC}_{AB}\big]+\Delta S^{\cC}_2 ,
\end{equation}
where $\rho^\cC_{AB}$ denotes the averaged state. We call the two terms on the right-hand side the neutral part and the charged part of the mutual information. In the expansion at small $x$ the neutral part contains the term $c\,x^2/128$ of the vacuum, which is present in any theory, and the contributions of the neutral operators, while the charged operators contribute only to $\Delta S_2^\cC$. For a group \eqref{eq:split} holds for the von Neumann entropies as well, with $\Delta S$ equal to the relative entropy $S(\rho_{AB}\,\|\,\rho^{G_A}_{AB})$ \cite{Casini:2019kex,Petz:1988suff}. For a general fusion category we define the neutral part by \eqref{eq:LM} with $Z$ replaced by $Z_{\Pi_\cC}$, so that \eqref{eq:split} holds by definition. It is then the R\'enyi-2 mutual information of the subtheory in which the operators on $A$ are required to be $\cC$-neutral, but we have not shown that it is the mutual information of a density matrix on the original Hilbert space. This question should be addressed with the conditional expectations of \cite{AliAhmad:2025faces,Benedetti:2024dku}.

The second interpretation concerns the decomposition of $\rho_{AB}$ with respect to the charge of $A$. For a $\mathbb{Z}_N$ symmetry we write $g_A=\sum_q\omega^qP_q$, where $P_q$ is the projector onto charge $q$ in $A$ and $\omega=e^{2\pi i/N}$. Then
\[
\tr\big(\rho_{AB}\,g_A\rho_{AB}g_A^{-1}\big)=\sum_{q,q'}\omega^{q-q'}\tr\big(P_q\rho_{AB}P_{q'}\rho_{AB}\big),
\]
and each term $\tr(P_q\rho_{AB}P_{q'}\rho_{AB})$ is the squared Hilbert--Schmidt norm $\|P_q\rho_{AB}P_{q'}\|_2^2$ of the block of $\rho_{AB}$ between the sectors $q$ and $q'$, where $\|X\|_2=(\tr X^\dagger X)^{1/2}$. The average $\mathcal T_A\rho_{AB}=\sum_qP_q\rho_{AB}P_q$ is the part of $\rho_{AB}$ that is diagonal in the charge of $A$, and $\tr(\rho_{AB}\,\mathcal T_A\rho_{AB})=\tr(\mathcal T_A\rho_{AB})^2$. For $\mathbb{Z}_2$ it follows that
\begin{equation}\label{eq:HS}
\frac{\|\rho_{AB}-\mathcal T_A\rho_{AB}\|_2^2}{\tr\rho_{AB}^2}=1-e^{-\Delta S_2}=\frac{1-\cO_g}{2},
\end{equation}
so that $(1-\cO_g)/2$ is the fraction of $\tr\rho_{AB}^2$ that comes from the blocks of $\rho_{AB}$ off-diagonal in the charge of $A$. The relative asymmetry thus measures the extent to which the charge of $A$ fails to be a good quantum number in the state $\rho_{AB}$. It vanishes for distant intervals, and by \eqref{eq:bound} it is at most $\log\dim\cC$.\footnote{One may also consider an interval $A$ separated from a region $B$ by a buffer $C$ on both sides, and the conditional mutual information $I(A{:}B\,|\,C)$. In the ground state on the line $A\cup B\cup C$ is the whole system and the state is pure, so that $I(A{:}B\,|\,C)=I(A{:}B)$ for any R\'enyi index. Moreover $B$ is then a single interval that contains the point at infinity. The configuration is therefore the one studied here, with cross-ratio $1-x_C$, where $x_C$ is the cross-ratio of the two components of $C$.}

\section{The $(E_8)_1$ theory}\label{sec:E8}

\subsection{The theory and its lines}

The chiral algebra of the $(E_8)_1$ WZW model is generated by 248 currents of weight one. It has a single irreducible representation, the vacuum module, with character
\[
\chi_{E_8}(\tau)=\frac{E_4(\tau)}{\eta(\tau)^8}=\fq^{-1/3}\big(1+248\fq+4124\fq^2+\dots\big)=j(\tau)^{1/3},
\]
where $E_4$ is the Eisenstein series of weight four, $\eta$ the Dedekind eta function and $j$ the modular invariant. The theory has $c=8$ and partition function $Z=|\chi_{E_8}|^2$. On the replica torus $j=256(1-x+x^2)^3/(x^2(1-x)^2)$.

We consider lines that act only on the holomorphic sector, and we obtain them as in \cite{Umasankar:2026tdl}. Let $V$ and $\tilde V$ be two subalgebras of the $E_8$ current algebra, each of which is the commutant of the other, such that the vacuum character decomposes as $\chi_{E_8}=\sum_{ij}M_{ij}\chi_i\tilde\chi_j$ in terms of their characters. A Verlinde line of $V$ labelled by a primary $a$ acts on the representation $i$ of $V$ as multiplication by $S_{ai}/S_{0i}$, where $S$ is the modular matrix of $V$. Acting in this way on the first factor of each term of the decomposition defines a topological line of $(E_8)_1$. It commutes with the stress tensor of $E_8$, which is the sum of those of $V$ and $\tilde V$. The corresponding twined character is
\[
\chi_L=\sum_{ij}\frac{S_{ai}}{S_{0i}}M_{ij}\chi_i\tilde\chi_j .
\]
The antiholomorphic character is not affected and cancels in \eqref{eq:obs}, so that $\cR_L$ is a function of $x$ and not of $\bar x$. The two invertible lines considered below do not depend on this construction. They are elements of order two and three of the group $E_8$, whose centralizers have the Lie algebras $D_8$ and $A_2\oplus E_6$, and are therefore symmetries of the theory. For the Fibonacci line we rely on the construction and on \cite{Albert:2025haag}, and the integrality of its defect partition function is a consistency check. The data of the three lines considered below are collected in table~\ref{tab:lines}, and the characters that enter are given in appendix~\ref{app:data}.

\subsection{Mutual information}

Before turning to the lines we record the mutual information. For real $x$ the character is real, so that $Z=|\chi_{E_8}|^2=\chi_{E_8}^2=j^{2/3}$. Substituting this in \eqref{eq:LM}, the powers of $2$, $x$ and $1-x$ combine to give
\begin{equation}\label{eq:I2}
I_2^{E_8}(x)=2\log\frac{1-x+x^2}{1-x}=2x^2+2x^3+x^4+O(x^6).
\end{equation}
The term of order $x^2$ receives a contribution $c/128=\frac1{16}$ from the vacuum and $2\cdot248/256=\frac{31}{16}$ from the currents. For $x\to1$ one finds $I_2\simeq\frac c4\log\frac1{1-x}$, as in any conformal field theory. The twist-field correlator \eqref{eq:G} is $G=|(1-x+x^2)/(x(1-x))|^2$.

\subsection{A $\mathbb{Z}_2$ line}\label{sec:z2line}

The first line generates a $\mathbb{Z}_2$ symmetry under which the 248 currents split into 120 even ones, which form the subalgebra $D_8=\mathfrak{so}(16)$, and 128 odd ones. In the construction described above it arises from the pairs $(B_{1,1},B_{6,1})$ and $(D_{2,1},D_{6,1})$ \cite{Umasankar:2026tdl}. Its twined character is $(\eta(\tau)/\eta(2\tau))^8=\fq^{-1/3}(1-8\fq+\dots)$, and expressing $\eta$ and $E_4$ through theta functions one finds
\begin{equation}\label{eq:RZ2}
\cR_{\mathbb{Z}_2}(x)=\frac{1-x}{1-x+x^2}=\frac{1}{1+2^8\big(\eta(2\tau)/\eta(\tau)\big)^{24}} .
\end{equation}
According to the second expression, $\cR_{\mathbb{Z}_2}$ is a rational function of $t_2=(\eta(\tau)/\eta(2\tau))^{24}$. This function is a Hauptmodul for $\Gamma_0(2)$, the group of modular transformations whose lower-left entry is even. The modular functions for $\Gamma_0(2)$ are the rational functions of $t_2$, just as those for the full modular group are the rational functions of $j$.

The defect Hilbert space has character $16(\eta(\tau)/\eta(\tau/2))^8=\fq^{1/6}(16+128\fq^{1/2}+\dots)$. Its lowest states have $h=\frac12$ and form the vector representation of $\mathfrak{so}(16)$. All weights are integers or half-integers, in agreement with the fact that \eqref{eq:RZ2} is single valued, and the behaviour $\cR_{\mathbb{Z}_2}\simeq1-x$ for $x\to1$ is \eqref{eq:large} with $n_L=16$ and $\Delta_L=\frac12$. From \eqref{eq:dS},
\begin{equation}\label{eq:dSZ2}
\Delta S_2^{\mathbb{Z}_2}(x)=\log\frac{2(1-x+x^2)}{2-2x+x^2},
\end{equation}
which behaves as $x^2/2$ for small $x$, equals $\log\frac65$ at $x=\frac12$, and approaches $\log2-(1-x)$ for $x\to1$.

\subsection{A $\mathbb{Z}_3$ line}

The second line generates a $\mathbb{Z}_3$ symmetry whose invariant currents form $A_2\oplus E_6$, of dimension $8+78$. It arises from the pair $(A_{2,1},E_{6,1})$ and from the decomposition with respect to $A_{2,1}^{\otimes4}$ associated with the tetracode \cite{Umasankar:2026tdl}. Let $A$ and $C$ be the theta functions of the $A_2$ lattice and of its translate by a fundamental weight, which are given in appendix~\ref{app:data}. Then $\chi_{E_8}=(A^4+8AC^3)/\eta^8$ and $\chi_{\mathbb{Z}_3}=(A^4-AC^3)/\eta^8$, and with the cubic identity of \cite{Borwein:1991cubic} we obtain
\begin{equation}\label{eq:RZ3}
\cR_{\mathbb{Z}_3}=\frac{A^3-C^3}{A^3+8C^3}=\frac{1}{1+3^5\big(\eta(3\tau)/\eta(\tau)\big)^{12}}.
\end{equation}
Here $(\eta(\tau)/\eta(3\tau))^{12}$ is a Hauptmodul for $\Gamma_0(3)$. As a function of $x$, $\cR_{\mathbb{Z}_3}$ satisfies the quartic equation
\[
j\,\cR^3(\cR-1)+27(1+8\cR)=0 ,
\]
with $j$ as given above. At $x=\frac12$, where $j=1728$, the relevant root is $\cR_{\mathbb{Z}_3}=\frac{1+\sqrt3}{4}$, and $\Delta S_2^{\mathbb{Z}_3}=\log(3-\sqrt3)$.

The defect Hilbert space is described in appendix~\ref{app:data}. Its lowest states have $h=\frac13$ and degeneracy three, and all its weights are multiples of $\frac13$. Correspondingly $\cR_{\mathbb{Z}_3}\simeq3((1-x)/16)^{2/3}$ for $x\to1$, and there is a branch point of order three at $x=1$, as expected from section~\ref{sec:limits}.

\subsection{The Fibonacci line}

The third line is not invertible. Under the subalgebra $(G_2)_1\times(F_4)_1$ the vacuum character decomposes as $\chi_{E_8}=\chi_0\tilde\chi_0+\chi_{2/5}\tilde\chi_{3/5}$, where the subscripts are the conformal weights. Each of the two factors has two primaries, with Fibonacci fusion rules \cite{Mathur:1988na}. Let $W$ be the non-trivial Verlinde line of $(G_2)_1$. It acts as $\varphi$ on the vacuum module of $(G_2)_1$ and as $-\varphi^{-1}$ on the other module. The twined character, and the character of the defect Hilbert space that follows from it by the transformation $S$, are
\begin{equation}\label{eq:ZW}
\chi_W=\varphi\,\chi_0\tilde\chi_0-\varphi^{-1}\chi_{2/5}\tilde\chi_{3/5},\qquad
\chi^W=\chi_0\tilde\chi_{3/5}+\chi_{2/5}\tilde\chi_0+\chi_{2/5}\tilde\chi_{3/5}.
\end{equation}
The coefficients in $\chi^W$ are non-negative integers, and the weights are in $\{0,\frac25,\frac35\}+\mathbb{Z}$, as required for a line with Fibonacci fusion rules \cite{Chang:2018iay}. This is consistent with the Fibonacci symmetry of $(E_8)_1$ discussed in \cite{Albert:2025haag}. The lowest states of $\cH_W$ have $h=\frac25$ and transform in the seven-dimensional representation of $G_2$, so that $n_W=7$. The relative asymmetry for the category generated by $W$ approaches $\log(1+\varphi^2)=\log(2+\varphi)\approx1.286$ for $x\to1$.

One can check numerically that $\cR_W$ does not satisfy the analogue of \eqref{eq:RZ3} with a Hauptmodul of $\Gamma_0(5)$. A closed expression can nevertheless be obtained as follows. The characters of a rational theory with two primaries are hypergeometric functions of $1728/j$ \cite{Mathur:1988na,Naculich:1988xv}. If the central charge is $c$ and the non-trivial primary has weight $h$ and a leading coefficient $N$,
\begin{equation}\label{eq:hyper}
\chi_0=j^{\frac c{24}}\,{}_2F_1\!\Big(-\tfrac c{24},\tfrac13-\tfrac c{24};1-h;\tfrac{1728}j\Big),\qquad
\chi_h=N\,j^{\frac c{24}-h}\,{}_2F_1\!\Big(h-\tfrac c{24},h+\tfrac13-\tfrac c{24};1+h;\tfrac{1728}j\Big).
\end{equation}
For $(G_2)_1$ one has $(c,h,N)=(\frac{14}5,\frac25,7)$ and for $(F_4)_1$ one has $(\frac{26}5,\frac35,26)$. In terms of the cross-ratio, $1728/j=27x^2(1-x)^2/(4(1-x+x^2)^3)$.

\begin{figure}[!htb]
\centering
\includegraphics[width=0.9\textwidth]{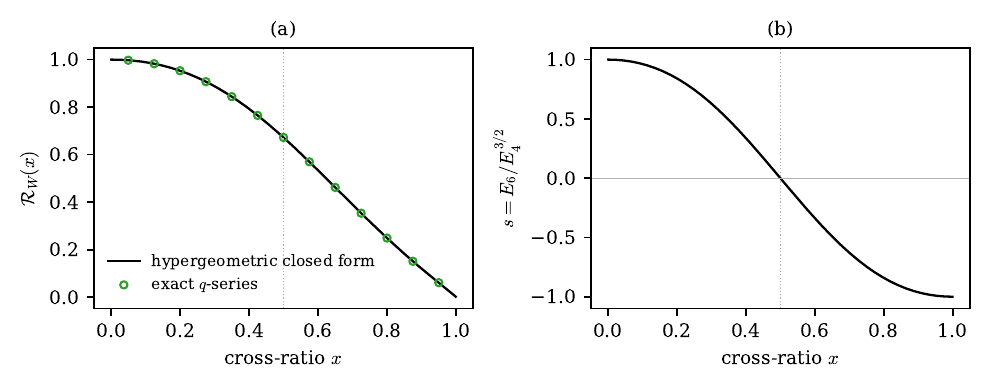}
\caption{(a) The Fibonacci overlap $\cR_W$ computed from the hypergeometric expression (line) and from the $\fq$-expansion of the characters (circles). (b) The variable $s=E_6/E_4^{3/2}$ as a function of $x$. It vanishes at $x=\frac12$ and is odd under $x\to1-x$.}
\label{fig:fib}
\end{figure}

These formulas cannot be used as they stand for all $x$. Indeed $j$ is invariant under $x\to1-x$, whereas $\cR_W$ is not. The reason is that $1728/j$ reaches the value one at $x=\frac12$, which corresponds to $\tau=i$ and is a square-root branch point of the hypergeometric functions, and the characters for $x>\frac12$ are given by the other branch. It is therefore convenient to introduce
\begin{equation}\label{eq:sunif}
s=\frac{E_6}{E_4^{3/2}}=\frac{(1+x)(2-x)(1-2x)}{2(1-x+x^2)^{3/2}},\qquad s^2=1-\frac{1728}j ,
\end{equation}
which decreases from $1$ to $-1$ as $x$ goes from $0$ to $1$ and changes sign under $x\to1-x$. The hypergeometric functions in \eqref{eq:hyper} are of the form ${}_2F_1(a,b;\gamma;1-s^2)$ with $\gamma-a-b=\frac12$, and for $0<s<1$ the standard connection formula gives
\[
{}_2F_1\big(a,b;\gamma;1-s^2\big)=\frac{\Gamma(\gamma)\Gamma(\frac12)}{\Gamma(\gamma-a)\Gamma(\gamma-b)}\,{}_2F_1\big(a,b;\tfrac12;s^2\big)+\frac{\Gamma(\gamma)\Gamma(-\frac12)}{\Gamma(a)\Gamma(b)}\,s\,{}_2F_1\big(\gamma-a,\gamma-b;\tfrac32;s^2\big).
\]
The right-hand side is analytic at $s=0$ and provides the continuation to negative $s$. Inserting it in \eqref{eq:hyper}, we obtain $\cR_W=(\chi_0\tilde\chi_0-\varphi^{-2}\chi_{2/5}\tilde\chi_{3/5})/(\chi_0\tilde\chi_0+\chi_{2/5}\tilde\chi_{3/5})$ for all $0<x<1$. We have checked the result against the $\fq$-expansion of the characters (figure~\ref{fig:fib}).

At $x=\frac12$ only the first term of the connection formula contributes. If we write $\cR_W=(1-\varphi^{-2}T)/(1+T)$ with $T=\chi_{2/5}\tilde\chi_{3/5}/\chi_0\tilde\chi_0$, then
\[
T\big(\tfrac12\big)=\frac{91}{864}\,\frac{\Gamma(\frac75)\Gamma(\frac85)\Gamma(\frac{17}{60})\Gamma(\frac{23}{60})\Gamma(\frac{37}{60})\Gamma(\frac{43}{60})}{\Gamma(\frac25)\Gamma(\frac35)\Gamma(\frac{47}{60})\Gamma(\frac{53}{60})\Gamma(\frac{67}{60})\Gamma(\frac{73}{60})},
\]
which gives $\cR_W(\frac12)\approx0.672$. The same method applies to any theory with two characters.

\begin{table}[!htb]
\centering
\renewcommand{\arraystretch}{1.3}
\begin{tabular}{lccccccc}
\toprule
line & fixed currents & $d_L$ & $\dim\cC$ & $\Delta_L$ & $n_L$ & $\cR_L(\tfrac12)$ & $\Delta S_2^\cC(\tfrac12)$\\
\midrule
$\mathbb{Z}_2$ & $D_8$ \ (120) & $1$ & $2$ & $1/2$ & $16$ & $2/3$ & $\log\tfrac65$\\
$\mathbb{Z}_3$ & $A_2E_6$ \ (86) & $1$ & $3$ & $1/3$ & $3$ & $\tfrac{1+\sqrt3}{4}$ & $\log(3-\sqrt3)$\\
$W$ & $G_2F_4$ \ (66) & $\varphi$ & $2+\varphi$ & $2/5$ & $7$ & $0.67229$ & $0.27067$\\
\bottomrule
\end{tabular}
\caption{Data of the three lines of $(E_8)_1$. The second column gives the subalgebra of currents on which $\hat L$ acts as $d_L$ and its dimension, which determines $N_\perp=128$, $162$ and $182$ in \eqref{eq:dSasym}. $\Delta_L$ and $n_L$ are the lowest dimension in $\cH_L$ and its degeneracy. The spins of the states in $\cH_L$ are in $\frac12\mathbb{Z}$, $\frac13\mathbb{Z}$ and $\{0,\frac25,\frac35\}+\mathbb{Z}$ respectively. The last two columns are the values at $x=\frac12$, where $I_2=2\log\frac32$.}
\label{tab:lines}
\end{table}

\begin{figure}[!htb]
\centering
\includegraphics[width=\textwidth]{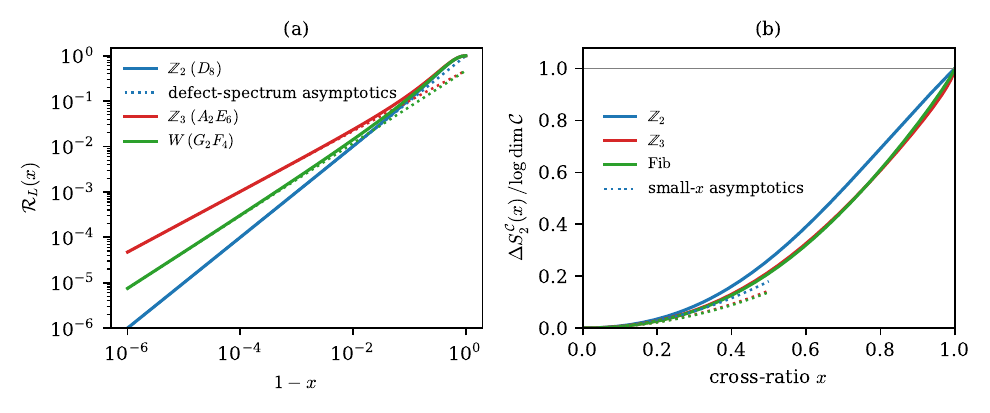}
\caption{(a) The normalized overlaps $\cR_L$ for the three lines of $(E_8)_1$ as functions of $1-x$, on logarithmic scales. The dotted lines are the asymptotic form \eqref{eq:large}, with slopes $2\Delta_L=1$, $\frac23$ and $\frac45$. (b) The relative asymmetry in units of $\log\dim\cC$. The dotted lines are the small-$x$ behaviour $N_\perp x^2/256$.}
\label{fig:ratios}
\end{figure}

The overlaps and relative asymmetries of the three lines are plotted in figure~\ref{fig:ratios}. As a check of the relation $Z_L(\tau)=Z^L(\tau')$, we computed for each line $Z_L$ at cross-ratio $x$ and $Z^L$ at $1-x$ from the $\fq$-expansions of the respective characters, and verified that they agree. For small $x$ the three curves in figure~\ref{fig:ratios}a are close to each other. By \eqref{eq:small}, $1-\cR_L\simeq(248-\tr_{\mathbf{248}}\hat L/d_L)\,x^2/256$, and the coefficient of $x^2$ is $1$, $243/256$ and approximately $0.98$ for the three lines.

\section{The critical Ising model}\label{sec:ising}

The lines of $(E_8)_1$ considered above are not easily realized in a lattice model. In order to compare with a lattice computation we turn to the Ising model, which has both an invertible and a non-invertible line.

\subsection{Exact results}

The Ising model has $c=\frac12$ and three primary fields, the identity, the energy operator $\varepsilon$ of weights $(\frac12,\frac12)$ and the spin field $\sigma$ of weights $(\frac1{16},\frac1{16})$. Its simple topological lines are the identity, the $\mathbb{Z}_2$ line $\eta$ and the Kramers--Wannier duality line $N$, with $\eta\times\eta=1$, $\eta\times N=N$ and $N\times N=1+\eta$ \cite{Frohlich:2006ch,Chang:2018iay}. On the three primaries $\eta$ has eigenvalues $(1,1,-1)$ and $N$ has eigenvalues $(\sqrt2,-\sqrt2,0)$. The three lines form the Ising fusion category, of dimension $1+1+2=4$. Since $N\times N$ contains $\eta$, there is no smaller category that contains $N$. The average \eqref{eq:dS} can thus be taken either over $\{1,\eta\}$ or over all three lines. In the first case the projector \eqref{eq:Pi} retains the modules of the identity and of $\varepsilon$. In the second it retains only the module of the identity, because $N$ acts on $\varepsilon$ as $-\sqrt2$.

The Virasoro characters are $\chi_0\pm\chi_{1/2}=\sqrt{\theta_{3,4}/\eta}$ and $\chi_{1/16}=\sqrt{\theta_2/2\eta}$. Since $\theta_2^4/\theta_3^4=x$ and $\theta_4^4/\theta_3^4=1-x$, the quantities of section~\ref{sec:loop} are algebraic functions of the cross-ratio. For $0<x<1$, with $u=x^{1/4}$ and $v=(1-x)^{1/4}$, we find
\begin{equation}\label{eq:isingI2}
I_2(x)=\log\frac{1+u+v}{2\sqrt v},\qquad
\cO_\eta(x)=\frac{1-u+v}{1+u+v},\qquad
\frac{\cO_N(x)}{\sqrt2}=\frac{2\sqrt v}{1+u+v}.
\end{equation}
The first expression is the known result for the mutual information \cite{Alba:2009ek,Alba:2011fu}. The other two are, by \eqref{eq:dS}, linear combinations of the partition functions of the Ising model and of its two subtheories, which were obtained in \cite{Benedetti:2026jones}. Here they appear as the overlaps of the individual lines, and one can verify the statements of section~\ref{sec:limits}. At small $x$, $1-\cO_\eta\simeq x^{1/4}$, which is the contribution of the spin field, with $\Delta=\frac18$ and $\lambda_\eta=-1$. For $x\to1$, $\cO_\eta\simeq((1-x)/16)^{1/4}$, corresponding to a single state of dimension $\frac18$ in $\cH_\eta$, the disorder operator. For the duality line $\cO_N\simeq2((1-x)/16)^{1/8}$, in agreement with the two states of weights $(\frac1{16},0)$ and $(0,\frac1{16})$ in $\cH_N$.

The lines of the Ising model act on both chiral halves, and for complex $x$ the overlaps depend on $x$ and $\bar x$. In \eqref{eq:isingI2} one should then replace $u$ and $v$ by $|x|^{1/4}$ and $|1-x|^{1/4}$, and $\sqrt v$ in the numerator of $\cO_N$ by the real part of $(1-x)^{1/8}$. Thus $\cO_\eta$ is single valued around $x=1$, as it must be since the spins in $\cH_\eta$ are integers or half-integers, whereas $\cO_N$ has a branch point. In the notation of \eqref{eq:monodromy} it is the sum of two eigenfunctions, proportional to $(1-x)^{1/8}$ and to its complex conjugate, with eigenvalues $e^{\pm i\pi/4}$, which reflects the spins $\pm\frac1{16}$ modulo $\frac12$ of the states in $\cH_N$.

The relative asymmetries for the two choices of category are
\begin{equation}\label{eq:isingdS}
\Delta S_2^{\mathbb{Z}_2}(x)=\log\frac{1+u+v}{1+v},\qquad
\Delta S_2^{\rm Ising}(x)=\log\frac{2(1+u+v)}{(1+\sqrt v)^2}.
\end{equation}
They increase from zero to $\log2$ and to $\log4$ respectively. In the language of section~\ref{sec:qi} they are the charged parts of the mutual information. Since the lightest operator of the Ising model is $\sigma$, which is charged, the mutual information is almost entirely charged except close to $x=1$ (figure~\ref{fig:qi}a). At $x=\frac12$, for example, $I_2=0.380$, of which $0.376$ is charged with respect to $\mathbb{Z}_2$ and $0.378$ with respect to the full category.

\subsection{A lattice computation}\label{sec:lattice}

The transverse-field Ising chain
\[
H=-\sum_j\sigma^x_j\sigma^x_{j+1}-\sum_j\sigma^z_j
\]
is critical and is described at long distances by the Ising conformal field theory. Its $\mathbb{Z}_2$ symmetry is generated by $\prod_j\sigma^z_j$, and the restriction of the generator to a block $A$ of sites is $P_A=\prod_{j\in A}\sigma^z_j$. The lattice counterpart of $\cO_\eta$ is therefore
\[
\frac{\tr(\rho_{AB}P_A\rho_{AB}P_A)}{\tr\rho_{AB}^2}.
\]
We evaluated this ratio in the ground state of the infinite chain for two blocks of $\ell$ sites separated by $d$ sites, for which $x=\ell^2/(\ell+d)^2$. The computation is exact and does not use tensor networks. It relies on the representation of the chain in terms of free fermions. After the Jordan--Wigner transformation the ground state is a Gaussian state of Majorana fermions, with correlation matrix elements $2/(\pi(2r+1))$ at separation $r$. The reduced density matrix of two disjoint blocks of spins is not Gaussian, because the Jordan--Wigner string of the second block passes through the sites between the blocks, but it is a linear combination of four Gaussian operators \cite{Fagotti:2010yr}. One of them is constructed from the fermionic correlation matrix of $A\cup B$. A second one is constructed from the Schur complement obtained by integrating out the intermediate sites, and is weighted by the expectation value of the fermion parity of those sites. The other two are the conjugates of these by the fermion parity of $B$. Since $P_A$ is the fermion parity of $A$, which is also Gaussian, the numerator and the denominator reduce to sums of Pfaffians of matrices of size $8\ell$. The cost grows only polynomially with $\ell$, which is what makes blocks of more than a hundred sites accessible. For small blocks we checked the procedure against a matrix-product-state computation.

\begin{figure}[!htb]
\centering
\includegraphics[width=\textwidth]{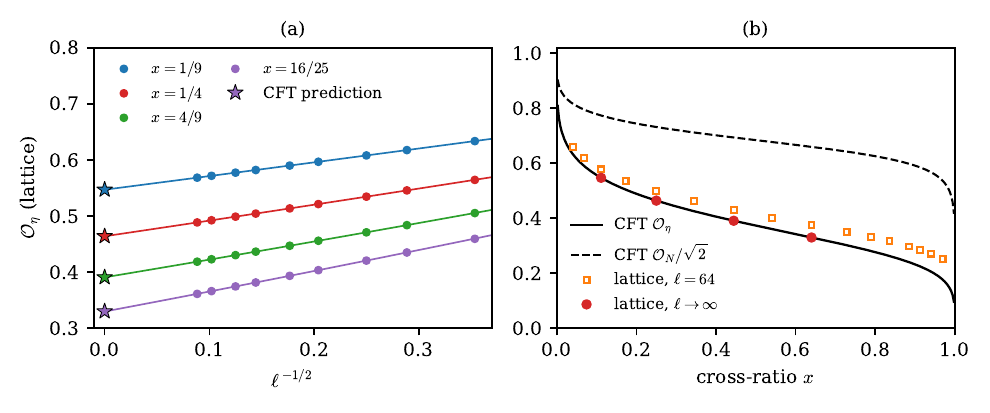}
\caption{The overlap $\cO_\eta$ in the critical transverse-field Ising chain. (a) Lattice values for two blocks of $\ell$ sites at four values of $x$, as functions of $\ell^{-1/2}$. The curves are fits to a constant plus terms in $\ell^{-1/2}$ and $\ell^{-1}$, and the stars are the values of \eqref{eq:isingI2}. (b) $\cO_\eta$ (solid) and $\cO_N/\sqrt2$ (dashed) from \eqref{eq:isingI2}, with the lattice data for $\ell=64$ (open squares) and the extrapolated values (filled circles).}
\label{fig:ising}
\end{figure}

\begin{figure}[!htb]
\centering
\includegraphics[width=\textwidth]{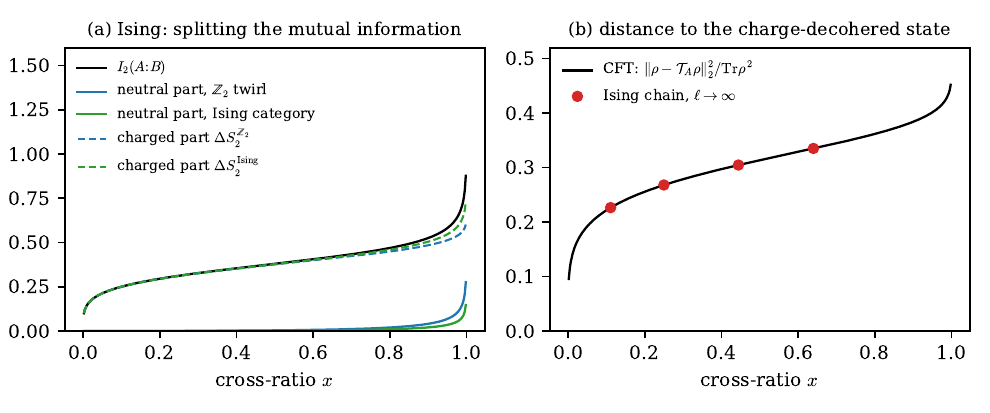}
\caption{(a) The R\'enyi-2 mutual information of two intervals in the Ising model and its neutral and charged parts \eqref{eq:split}, for $\mathbb{Z}_2$ and for the Ising fusion category. (b) The ratio \eqref{eq:HS} in the Ising model, with the extrapolated lattice values.}
\label{fig:qi}
\end{figure}

The results are shown in figure~\ref{fig:ising} and table~\ref{tab:ising}. At fixed $x$ the lattice values vary with $\ell$ as $\ell^{-1/2}$ to leading order. Corrections to scaling with exponent $1/n$ are known for the R\'enyi entropies in this chain \cite{Cardy:2010zs,Fagotti:2010yr}. We fitted the data for $16\le\ell\le128$ to a constant plus terms in $\ell^{-1/2}$ and $\ell^{-1}$. The extrapolated values agree with \eqref{eq:isingI2} for the four values of $x$ that we considered, and the same procedure applied to the mutual information reproduces the first formula in \eqref{eq:isingI2}. The agreement is stable under changes of the range of $\ell$ and of the number of terms in the fit (appendix~\ref{app:lattice}). By \eqref{eq:HS} the same data determine the fraction of $\tr\rho_{AB}^2$ carried by the blocks of $\rho_{AB}$ that are off-diagonal in the charge of $A$, which in the continuum is $x^{1/4}/(1+x^{1/4}+(1-x)^{1/4})$ (figure~\ref{fig:qi}b). We have not attempted a lattice computation of $\cO_N$. It would require the lattice form of the duality defect, which is non-invertible and mixes with lattice translations \cite{Seiberg:2023cdc}.

\begin{table}[!htb]
\centering
\renewcommand{\arraystretch}{1.2}
\begin{tabular}{ccccccc}
\toprule
& \multicolumn{3}{c}{$\mathcal{O}_\eta$} & \multicolumn{3}{c}{$I_2$}\\
\cmidrule(lr){2-4}\cmidrule(lr){5-7}
$x$ & $\ell=128$ & extrapolated & CFT & $\ell=128$ & extrapolated & CFT\\
\midrule
$1/9$ & 0.56827 & 0.54681 & 0.54688 & 0.24334 & 0.25705 & 0.25702\\
$1/4$ & 0.48842 & 0.46376 & 0.46385 & 0.29622 & 0.31277 & 0.31272\\
$4/9$ & 0.41840 & 0.39052 & 0.39064 & 0.34664 & 0.36612 & 0.36608\\
$16/25$ & 0.36103 & 0.32955 & 0.32977 & 0.39371 & 0.41625 & 0.41627\\
\bottomrule
\end{tabular}

\caption{Lattice values of $\cO_\eta$ and of the mutual information $I_2$ at $\ell=128$ and extrapolated from $16\le\ell\le128$, compared with \eqref{eq:isingI2}.}
\label{tab:ising}
\end{table}

\section{What the monodromy knows about the category}\label{sec:anom}

The overlaps computed so far are sensitive to the way in which each line acts on the spectrum, through \eqref{eq:small}, and to its defect spectrum, through \eqref{eq:large}. The fusion rules enter through the composition of loops, since two concentric loops around $A$ are equivalent to the sum of the loops of their fusion products. A fusion category contains more information than its fusion ring. It also has associativity data, the $F$-symbols, and two categories can share the same fusion rules and differ in $F$. In this section we show that part of this information is visible in $\cO_L$ through its monodromy around $x=1$. The selection rules on which this relies are known \cite{Chang:2018iay,Lin:2019kpn}. Our point is that they can be read off from the replica observable.

The key fact is that the spins of the states in a defect Hilbert space are not arbitrary. Going twice around the spatial circle in the presence of a line on the time cycle is the transformation $T^2$, and its action on $\cH_L$ is fixed by the category through its $F$-symbols \cite{Chang:2018iay}. For a non-anomalous $\mathbb{Z}_2$ the spins in $\cH_g$ are integers or half-integers. For a $\mathbb{Z}_N$ symmetry with 't~Hooft anomaly $k\in\mathbb{Z}_N$ the spins in the sector of the generator satisfy \cite{Chang:2018iay,Lin:2019kpn}
\begin{equation}\label{eq:spinZN}
s\in\frac{k}{N^2}+\frac{\mathbb{Z}}{N}.
\end{equation}
By \eqref{eq:monodromy}, the eigenvalues of the monodromy of $\cO_g$ around $x=1$ are the phases $e^{4\pi is}$. They form the set $e^{4\pi ik/N^2}\,\omega^{2\mathbb{Z}}$ with $\omega=e^{2\pi i/N}$, and raising them to the power $N$ for odd $N$, or $N/2$ for even $N$, removes the ambiguity and leaves $e^{4\pi ik/N}$ or $e^{2\pi ik/N}$. In either case the anomaly $k$ is determined by the eigenvalues. For $\mathbb{Z}_2$ the statement is particularly simple. If the symmetry is anomalous every state in $\cH_g$ has $s\in\pm\frac14+\mathbb{Z}$, so that $e^{4\pi is}=-1$ for all of them. There is then a single eigenvalue, and $\cO_g$ itself changes sign under the monodromy.

The standard example of an anomalous $\mathbb{Z}_2$ is the centre of $SU(2)$ in the $SU(2)_1$ WZW model. Since an anomalous symmetry cannot be made on-site, $\cO_g$ is defined here by the loop on the replica torus, as explained in section~\ref{sec:loop}. The theory has the two characters $\chi_0=\theta_3(2\tau)/\eta(\tau)$ and $\chi_{1/2}=\theta_2(2\tau)/\eta(\tau)$, the centre acts as $-1$ on the module of spin $\frac12$, and the defect partition function $\chi_0\bar\chi_{1/2}+\chi_{1/2}\bar\chi_0$ has lowest states of weights $(\frac14,0)$ and $(0,\frac14)$, each with multiplicity two. For real $x$, the identities $\theta_3(2\tau)^2\pm\theta_2(2\tau)^2=\theta_{3,4}(\tau)^2$ give
\begin{equation}\label{eq:su21}
\cO_g=\frac{\chi_0^2-\chi_{1/2}^2}{\chi_0^2+\chi_{1/2}^2}=\frac{\theta_4(\tau)^2}{\theta_3(\tau)^2}=\sqrt{1-x}.
\end{equation}
For complex $x$ one finds $\cO_g=2\operatorname{Re}r/(1+|r|^2)$ with $r=\theta_4(\tau/2)/\theta_3(\tau/2)\simeq\frac12\sqrt{1-x}$, and the square root changes sign around $x=1$. In the Ising model, by contrast, $\cO_\eta$ is single valued, as we saw in section~\ref{sec:ising}. The two $\mathbb{Z}_2$ symmetries have the same fusion rules and are distinguished by the monodromy of the overlap.

The same mechanism separates categories that share their fusion rules. The fusion rules $\eta\times\eta=1$, $\eta\times N=N$, $N\times N=1+\eta$ are realized by two inequivalent fusion categories, the Tambara--Yamagami categories based on $\mathbb{Z}_2$ \cite{Tambara:1998vmj}, which differ in the sign $\kappa=\pm1$ of one $F$-symbol. The Verlinde lines of the Ising model realize $\kappa=+1$, and those of $SU(2)_2$, whose spin-$\frac12$ line is the duality line, realize $\kappa=-1$. Both have $d_N=\sqrt2$, and in both the duality line acts as $\sqrt2$, $-\sqrt2$ and $0$ on the vacuum, on the module of weight $\frac12$ and on the remaining module. The spins in $\cH_N$, however, lie in $\pm\frac1{16}+\frac12\mathbb{Z}$ for $\kappa=+1$ and in $\pm\frac3{16}+\frac12\mathbb{Z}$ for $\kappa=-1$ \cite{Chang:2018iay}.

Both theories are free fermions, the Ising model with one Majorana fermion and $SU(2)_2$ with three, and they belong to a family. For $k$ Majorana fermions with the diagonal sum over spin structures, which is the $Spin(k)_1$ WZW model, the characters of the vacuum and vector modules are $\chi_0\pm\chi_v=(\theta_{3,4}/\eta)^{k/2}$, and those of the spinor modules are proportional to $(\theta_2/\eta)^{k/2}$. For odd $k$ there is one spinor module, and the corresponding line is a duality line with $d=\sqrt2$. For even $k$ there are two spinor modules with the same character, and the corresponding lines are invertible. In both cases a spinor line acts as $d$ on the vacuum module and as $-d$ on the vector module, the spinor modules do not contribute to its twined partition function, and the computation of section~\ref{sec:ising} gives, with $u=x^{1/4}$ and $v=(1-x)^{1/4}$,
\begin{equation}\label{eq:TY}
\cR_k(x)=\frac{2\,v^{k/2}}{1+u^k+v^k}.
\end{equation}
For complex $x$ the factor $v^{k/2}$ is replaced by $\operatorname{Re}(1-x)^{k/8}$ and $u^k,v^k$ by $|x|^{k/4},|1-x|^{k/4}$. The overlap is therefore the sum of two eigenfunctions of the monodromy \eqref{eq:monodromy}, proportional to $(1-x)^{k/8}$ and to its complex conjugate, with eigenvalues $e^{\pm i\pi k/4}$. These are the values of $e^{4\pi is}$ for the spins $\pm\frac k{16}$ of the lowest states in the defect Hilbert space. The overlap itself is not multiplied by a phase. Starting from real $x$, for example, one turn around $x=1$ multiplies it by $\cos\frac{\pi k}{4}$.

For the Ising model the eigenvalues are $e^{\pm i\pi/4}$, and for $SU(2)_2$ they are $e^{\pm3i\pi/4}$. The exponents $\frac18$ and $\frac38$ with which the two overlaps vanish depend on the theory, since they are twice the lowest dimension in $\cH_N$. The eigenvalues do not. Any theory whose duality line realizes the Ising category gives an overlap with monodromy eigenvalues $e^{\pm i\pi/4}$, and any theory realizing the other category gives $e^{\pm3i\pi/4}$. More generally, the pair of eigenvalues $e^{\pm i\pi k/4}$ depends on $k$ only modulo $8$, and it is the same for $k$ and $8-k$. This is the periodicity familiar from the classification of these theories \cite{Kitaev:2005hzj}. Odd values of $k$ equal to $\pm1$ modulo $8$ give the eigenvalues of the Ising category, and those equal to $\pm3$ give the eigenvalues of the other. Among the even values, $k=8$ gives a single-valued overlap. For $k=4$ the eigenvalue is $-1$, which signals an anomalous $\mathbb{Z}_2$. In that case the theory is the product of two copies of $SU(2)_1$, the spinor line is the centre of one of them, and \eqref{eq:TY} reduces to $\sqrt{1-x}$, in agreement with \eqref{eq:su21}. For $k=2$ the eigenvalues are $\pm i$, and the spinor line generates a $\mathbb{Z}_4$ symmetry with an anomaly.

These monodromies have a three-dimensional interpretation. The Verlinde lines of a diagonal rational theory are labelled by the anyons of the three-dimensional topological field theory whose boundary carries the chiral algebra, which for a WZW model is Chern--Simons theory \cite{Witten:1988hf,Fuchs:2002cm}. In the examples above the monodromy eigenvalues are $\theta_a^{\pm2}$, where $\theta_a$ is the topological spin of the anyon $a$ that labels the line. The spinor lines of the $Spin(k)_1$ models correspond to the vortices of the sixteenfold way \cite{Kitaev:2005hzj}, with $\theta=e^{2\pi ik/16}$, which include the duality lines of the Ising model and of $SU(2)_2$ for $k=1$ and $k=3$. The anomalous $\mathbb{Z}_2$ of $SU(2)_1$ corresponds to the semion of $SU(2)_1\simeq U(1)_2$ Chern--Simons theory, with $\theta=i$. This is the topological order of the Kalmeyer--Laughlin chiral spin liquid \cite{Kalmeyer:1987vb}, whose edge is described by the $SU(2)_1$ WZW model. The transition out of this phase is described by the $U(1)_2$ Chern--Simons--matter CFT, whose conformal data are bounded and estimated in \cite{Umasankar:2026pv}. Two-interval entanglement in the edge theory thus detects the topological spins of the phase adjacent to that critical point.

The monodromy is an analytic continuation, but the spins it detects can also be measured directly. For two intervals on a constant-time slice the cross-ratio is real, $\bar x=x$, and only the scaling dimensions $\Delta=h+\bar h$ enter the overlaps. The monodromy argument requires complex $x$ with $\bar x=x^*$, which is a Euclidean configuration rather than a pair of intervals on a time slice. For intervals on a boosted spacelike slice the two light-cone cross-ratios $x$ and $\bar x$ are independent numbers in $(0,1)$, the moduli $\tau$ and $-\bar\tau$ are independent points on the imaginary axis, and the nearly adjacent limit becomes
\[
\cO_L(x,\bar x)\simeq\sum n_{h,\bar h}\Big(\frac{1-x}{16}\Big)^{2h}\Big(\frac{1-\bar x}{16}\Big)^{2\bar h},
\]
from which $h$ and $\bar h$, and hence the spins in $\cH_L$, can be read off separately. For the centre of $SU(2)_1$, for example, the leading terms are $2\big(\frac{1-x}{16}\big)^{1/2}+2\big(\frac{1-\bar x}{16}\big)^{1/2}$, which reduces to \eqref{eq:su21} at $\bar x=x$ and shows the spins $\pm\frac14$ explicitly.

Genus one does not see everything. The torus partition functions with lines, including those with lines on both cycles and junctions between them, form representations of the tube algebra of $\cC$, and the modular transformations act on them through the modular data of the Drinfeld centre of $\cC$. Modular data are known not to determine a modular category \cite{Mignard2017ModularCA}. We return to this point in section~\ref{sec:disc}.

\section{Crossing equations at twice the central charge}\label{sec:blocks}

\subsection{From torus to sphere}

The correlator \eqref{eq:G} of twist fields in the orbifold theory, whose central charge is $2c$, can be expanded in Virasoro conformal blocks. The block $F_H(x)$ is the contribution of a Virasoro primary of weight $H$ in the operator product of the first two fields, together with all its descendants, and the coefficients of the expansion are sums of squares of OPE coefficients, which are non-negative in a unitary theory. The equality of the expansions around $x=0$ and around $x=1$ is the crossing equation.

For four fields of equal weight $h_\sigma$ at central charge $C$, the blocks have the representation \cite{Zamolodchikov:1984eqp,Zamolodchikov:1987avt}
\begin{equation}\label{eq:zam}
F_H(x)=(16q)^{H-\frac{C-1}{24}}\,[x(1-x)]^{\frac{C-1}{24}-2h_\sigma}\,\theta_3(q)^{\frac{C-1}{2}-16h_\sigma}\,\mathbf H_H(q),
\end{equation}
where $q$ is the nome introduced in section~\ref{sec:replica} and $\mathbf H_H(q)=1+O(q^2)$ is a power series in $q^2$ determined by a recursion relation. The three prefactors are the same for every block and contain the whole dependence on the external weights, while $\mathbf H_H$ carries the information about the exchanged representation. For the twist correlator $C=2c$ and $h_\sigma=c/16$, and the two exponents that involve $h_\sigma$ become
\[
\frac{C-1}{24}-2h_\sigma=-\frac{c}{24}-\frac1{24},\qquad\frac{C-1}{2}-16h_\sigma=-\frac12 .
\]
The first differs from the exponent $-c/24$ of $x(1-x)$ in \eqref{eq:G} by an amount that does not depend on $c$, and the second does not depend on $c$ at all. This is a property of the weight of the twist field, and it is the reason why the comparison with the torus is simple.

Since $Z$, $Z_L$ and $Z^L$ are sums of products of holomorphic and antiholomorphic characters, the expansion of $G$ in products $F_H(x)F_{\bar H}(\bar x)$ can be carried out separately for each holomorphic character, and we write the formulas for the holomorphic dependence only. Let $\chi_\bullet$ be a holomorphic character that appears in $Z$, $Z_L$ or $Z^L$. Equating $(2^8x(1-x))^{-c/24}\chi_\bullet(\tau)$, its contribution to \eqref{eq:G}, with $\sum_Ha^\bullet_HF_H(x)$, and moving the prefactors of \eqref{eq:zam} to the left, we obtain
\begin{equation}\label{eq:match}
\sum_Ha^\bullet_H\,(16q)^H\,\mathbf H_H(q)=2^{-c/3}\,(16q)^{\frac{2c-1}{24}}\,[x(1-x)]^{\frac1{24}}\,\theta_3^{1/2}\,\chi_\bullet(\tau).
\end{equation}
The right-hand side simplifies in three steps. In the first, the relations $x=\theta_2^4/\theta_3^4$ and $1-x=\theta_4^4/\theta_3^4$ give $x(1-x)=\theta_2^4\theta_4^4/\theta_3^8$, so that the factors that depend on $x$ combine into
\[
[x(1-x)]^{\frac1{24}}\,\theta_3^{1/2}=(\theta_2\theta_4)^{\frac16}\,\theta_3^{-\frac13+\frac12}=(\theta_2\theta_3\theta_4)^{\frac16}.
\]
In the second, the product of the three theta functions is $\theta_2\theta_3\theta_4=2\eta^3$, and $\eta=q^{1/12}\prod_n(1-q^{2n})$, so that
\[
(\theta_2\theta_3\theta_4)^{\frac16}=2^{\frac16}\,\eta^{\frac12}=2^{\frac16}\,q^{\frac1{24}}\prod_{n\ge1}(1-q^{2n})^{\frac12}.
\]
In the third, the remaining constants and powers of $q$ combine as
\[
2^{-\frac c3}\cdot16^{\frac{2c-1}{24}}\cdot2^{\frac16}=2^{-\frac c3+\frac{2c-1}6+\frac16}=1,\qquad q^{\frac{2c-1}{24}}\,q^{\frac1{24}}=q^{\frac c{12}}=\fq^{\frac c{24}}.
\]
The powers of $2$, including the factor $2^8$ in \eqref{eq:G}, cancel exactly, and \eqref{eq:match} becomes
\begin{equation}\label{eq:master}
\fq^{c/24}\prod_{n\ge1}(1-\fq^n)^{1/2}\,\chi_\bullet(\tau)=\sum_Ha^\bullet_H\,(16q)^H\,\mathbf H_H(q).
\end{equation}
The left-hand side is the character multiplied by a factor that is the same for every theory. A state of weight $h$ contributes $\fq^{h-c/24}$ to $\chi_\bullet$, and therefore $q^{2h}$ times a power series in $q^2$ to the left-hand side, since $\fq=q^2$. The right-hand side contains the powers $q^H$, again multiplied by power series in $q^2$. The exchanged weights are therefore of the form $H=2h+2k$, with $k$ a non-negative integer, and the coefficients $a^\bullet_H$ can be determined order by order in $q$. The relation between the twist-field correlator and the torus partition function is well known \cite{Lunin:2000yv,Headrick:2010zt}, and the recursion has been applied to two intervals before \cite{Ruggiero:2018hyl}. We use it here also for the sectors with a line. We computed $\mathbf H_H$ with the \texttt{virasoro\_blocks} package \cite{Poland:2026vb}, and we checked \eqref{eq:master} against direct evaluations of the blocks.

\subsection{Positive and bounded channels}\label{sec:channels}

\begin{figure}[!htb]
\centering
\includegraphics{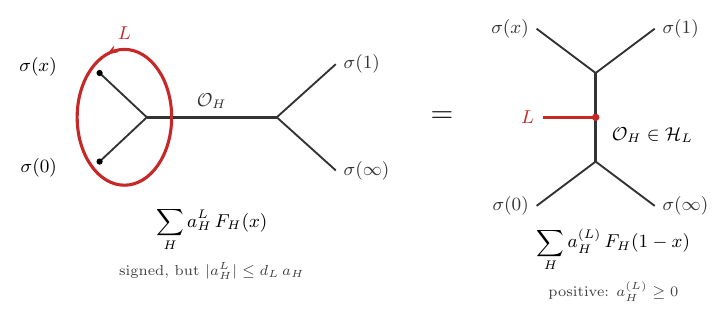}
\caption{The two sides of \eqref{eq:crossing}. On the left the line encircles the twist fields at $0$ and $x$ and acts on the exchanged states. On the right, in the crossed channel, the exchanged states belong to the defect Hilbert space.}
\label{fig:crossing}
\end{figure}

The coefficients in \eqref{eq:master} have the following properties, which are illustrated in figure~\ref{fig:crossing}. For the correlator without a line, $a_H\ge0$ by unitarity. If the line is on the $b$-cycle, the states exchanged between the twist fields at $0$ and $x$ and those at $1$ and $\infty$ belong to $\cH_L$ (figure~\ref{fig:torus}). The coefficients are again squared norms, and so $a_H^{(L)}\ge0$. If the line is on the $a$-cycle, it encircles the twist fields at $0$ and $x$ and acts on the exchanged states as $\hat L$ on one of the two copies. This operator commutes with the Virasoro algebra of the orbifold theory, and its eigenvalues are bounded in absolute value by $d_L$. Hence $|a^L_H|\le d_La_H$. The transformation $S$ relates the two configurations with a line,
\begin{equation}\label{eq:crossing}
\sum_Ha^L_HF_H(x)=\sum_Ha^{(L)}_HF_H(1-x).
\end{equation}
The coefficients on the left-hand side have no definite sign, but are bounded by those of the correlator without a line. The coefficients on the right-hand side are non-negative.

For $(E_8)_1$ we computed the coefficients up to $H=14$. They are listed in appendix~\ref{app:blocks}. All $a_H$ and $a_H^{(L)}$ are positive, and the first coefficients in the defect sectors are the multiplicities of the lowest levels of $\cH_L$. Apart from the vacuum, the ratio $|a^L_H|/(d_La_H)$ is largest at $H=2$, where it is equal to $\frac1{31}$, $\frac5{248}$ and $0.0142$ for the three lines. With the sums truncated at $H=14$ we verified numerically that the two sides of \eqref{eq:crossing} agree.

\subsection{Relation to the modular bootstrap}

The factor that relates $G$ to $Z$ in \eqref{eq:G} is invariant under $x\to1-x$. The crossing equation of the correlator without a line is therefore the modular invariance of $Z$, and \eqref{eq:crossing}, together with its antiholomorphic counterpart, is the relation $Z_L(\tau)=Z^L(-1/\tau)$. The block expansion adds to this the properties of the coefficients described in section~\ref{sec:channels}, and one may ask whether the resulting bootstrap problem differs from the modular bootstrap with symmetry lines \cite{Lin:2019kpn,Lin:2021udi}.

The answer depends on how the blocks are treated. The states of the orbifold theory descend from those of $T$, which are organized in representations of the Virasoro algebra of $T$. According to \eqref{eq:master}, a module of $T$ with $h>0$ contributes $q^{2h}\prod_n(1-q^{2n})^{-1/2}$ to the left-hand side, and the vacuum module contributes $(1-q^2)\prod_n(1-q^{2n})^{-1/2}$. Each module of $T$ therefore corresponds to a definite combination of blocks with $H=2h+2k$, $k=0,1,2,\dots$, whose coefficients we computed with \texttt{virasoro\_blocks} (table~\ref{tab:towers}). In all the cases we examined they are positive. For the vacuum module the block with $H=2$ is absent, whereas in the limit $h\to0$ of a generic module it appears with coefficient one in the normalization of the table. The difference is due to the null vector at level one in the vacuum module of $T$, and it cannot be seen if the blocks are treated as independent. The sum of the blocks associated with a module of $T$ is, up to the universal prefactor, the corresponding Virasoro character of $T$. If this structure is imposed, the crossing equations become the statement that $Z$, $Z_L$ and $Z^L$ are combinations of Virasoro characters related by $S$, with spins that are integers in $\cH$ and constrained by the category in $\cH_L$. This is the modular bootstrap with symmetry lines, without the relations among the defect sectors that follow from the transformations $T$ and $ST$. Conversely, since each module of $T$ corresponds to a positive combination of blocks, a functional that excludes a spectrum when the blocks are treated independently also excludes it when they are grouped into modules. The formulation with independent blocks is therefore the weaker of the two.

As an illustration we considered a non-anomalous $\mathbb{Z}_2$ symmetry at $c=8$. We wrote $a_H=a^+_H+a^-_H$ and $a^L_H=a^+_H-a^-_H$, with $a^\pm_H\ge0$ the contributions of even and odd states, and searched by linear programming for functionals built from derivatives of the two crossing equations at $z=\bar z=\frac12$, up to total order $7$ with independent blocks and $9$ with the blocks grouped into modules. Without assumptions on the spectrum no value of $\Delta_L$ can be excluded. This is to be expected. For a compact boson and the $\mathbb{Z}_2$ subgroup of its momentum symmetry, the operators on which the line ends have dimensions of order $R^2$ at large radius $R$, while the odd operators have dimensions of order $1/R^2$. If all odd scalars are assumed to have dimension larger than about $2.2$, we find $\Delta_L\lesssim0.6$, a value that is only approximate because of the discretization of the spectrum, and the two formulations give the same result. This is the phenomenon found in \cite{Lin:2019kpn}, where it was shown by the modular bootstrap that for a non-anomalous $\mathbb{Z}_2$ the lightest odd state and the lightest state of the defect Hilbert space cannot both be arbitrarily heavy. Their analysis includes more constraints and is carried out for general $c$, and we do not regard our numbers as new bounds. They are consistent with $(E_8)_1$, where for the $\mathbb{Z}_2$ line of section~\ref{sec:z2line} the lightest odd scalars have $\Delta=2$ and $\Delta_L=\frac12$. Whether the formulation in terms of a four-point function, in which the defect sector appears through the exchange of operators on which the line ends, has any practical advantage could only be decided by a treatment based on semidefinite programming that includes the $T$ and $ST$ relations.

\section{Discussion}\label{sec:disc}

We have studied the observable $\cO_L(x)$ defined by a topological line wrapped on one cycle of the replica torus of two intervals. For an on-site symmetry it is the overlap of the reduced density matrix of the two intervals with its image under the symmetry acting on one of them. Its average over a symmetry is the $n=2$ case of a known quantity, the difference of mutual informations between a theory and its neutral subtheory, and the bound \eqref{eq:bound} is the corresponding case of the known bounds in terms of $\log|G|$ \cite{Casini:2019kex} or of a Jones index \cite{Benedetti:2024dku,Benedetti:2026jones}. The results that we believe to be new concern the individual lines. As the intervals become adjacent, $\cO_L$ vanishes with an exponent equal to twice the lowest dimension in the defect Hilbert space of $L$, and the eigenvalues of its monodromy around $x=1$ are determined by the spins of the defect operators. Because these spins obey known selection rules fixed by the category, the observable detects 't~Hooft anomalies, as in $SU(2)_1$, and distinguishes categories with the same fusion rules, as for the duality lines of the Ising model and of $SU(2)_2$. These statements hold for any topological line of a unitary conformal field theory, whether or not it is invertible. The explicit expressions include non-invertible lines, the Fibonacci line of $(E_8)_1$ and the duality line of the Ising model, the latter being also obtainable from the subtheory partition functions of \cite{Benedetti:2026jones}, The comparison with the transverse-field Ising chain shows that, for an on-site symmetry, $\cO_L$ can be computed directly on the lattice. The crossing equations of section~\ref{sec:blocks} are a twist-field form of the modular bootstrap with symmetry lines, and we did not obtain new bounds from them.

A natural question is whether the full fusion category, with its fusion rules, quantum dimensions and $F$-symbols up to gauge transformations, can be reconstructed from entanglement data. Section~\ref{sec:anom} shows that genus one already detects part of the associativity data, and it is natural to go to higher genus. For $N$ intervals the replica surface of the $n$-th R\'enyi entropy has genus $(N-1)(n-1)$ \cite{Headrick:2015gba}, so genus two arises in two ways. Two intervals at $n=3$ give a $\mathbb{Z}_3$-symmetric cover with one modulus. Three intervals at $n=2$ give a hyperelliptic surface with three moduli, and since every surface of genus two is hyperelliptic, these are generic genus-two surfaces. Moreover, by the theorem of \cite{Birman1973Isotopies}, the mapping class group of genus two is generated by braiding the six branch points, so analytic continuation of three-interval R\'enyi-2 data in the cross-ratios realizes the full genus-two modular action. There is evidence that this goes beyond genus one. Modular data do not determine a modular category \cite{Mignard2017ModularCA}, but the known counterexamples are distinguished by representations of the genus-two mapping class group \cite{Wen:2019ylt}. Two further points would have to be addressed. The Drinfeld centre determines $\cC$ only up to Morita equivalence. The overlaps \eqref{eq:obs} also record how lines act on local operators, and it is conceivable that this additional information identifies $\cC$ itself, but we have not shown this. And networks with junctions require lines that cross the branch cuts, which correspond to density matrices of states dressed by defect strings, such as the disorder strings of the Ising chain. A concrete first test would be to recover the sign $\kappa$ of the Tambara--Yamagami categories, and then the $F$-symbols of the Fibonacci category, from genus-two data.

Another direction is the Haagerup fusion category $\mathcal H_3$, which is not known to arise from finite groups or from affine Lie algebras. Numerical studies of anyonic chains with this symmetry find critical points with $c\simeq2$ \cite{Huang:2021nvb,Vanhove:2021zop}, and it has been proposed in \cite{Albert:2025haag} that $(E_8)_1$ has lines forming $\mathcal H_3$, in connection with a theory with $c=8$ conjectured in \cite{Evans:2010yr}. For such lines the limiting value $\log\dim\mathcal H_3=\log\frac{39+9\sqrt{13}}{2}\approx3.576$ of the relative asymmetry is fixed by the category and does not test the proposal. The overlaps of the individual lines contain more information. In particular, the defect characters obtained from them by $S$ must have non-negative integer coefficients and spins compatible with $\mathcal H_3$. It would be interesting to compute them for $(E_8)_1$ with the methods of \cite{Umasankar:2026tdl}. In the anyonic chains one could measure $\cO_L$ by acting with the lattice symmetry operator on one block, and extract $\Delta_L$ from its behaviour as the blocks approach each other.

The analysis can be repeated for any theory in which the twined characters are known. An example is the Monster theory with its duality defect \cite{Lin:2019hks}, for which $c=24$ and the twist fields have weight $\frac32$. For $n>2$ the replica surface has genus $n-1$, and lines can be placed on several cycles. The derivation of \eqref{eq:bound} used only positivity in the dual channel, and we expect a similar bound to hold. Finally, under $\tau\to\tau+1$ the cross-ratio transforms as $x\to x/(x-1)$. The modular orbits of defect partition functions considered in \cite{Umasankar:2026tdl} therefore correspond to analytic continuations of $\cO_L$ in $x$, which we have not explored.

\acknowledgments
We thank David Poland for the \texttt{virasoro\_blocks} package and discussions. We are grateful to Arpit Das for the collaboration on commutant pairs and defect lines in $(E_8)_1$ on which the present work builds, and for his comments on symmetry-resolved entanglement. We thank Shankar Ganesh for suggesting the question of reconstructing the symmetry category from entanglement data, which led to section~\ref{sec:anom} and for comments on the draft. We thank Colin Coane and Daniel Qenani for comments on the draft and for stimulating discussions on quantum information, in particular on conditional mutual information and recovery maps, which prompted section~\ref{sec:qi}. We also thank Nayan Myerson-Jain and Wei Li for discussions. The numerical computations were performed with \texttt{virasoro\_blocks}, \texttt{TeNPy} and \texttt{pfapack}; we thank their authors. This work was supported by DOE grant DE-SC0017660.

\appendix
\section{The replica partition function}\label{app:replica}

We derive \eqref{eq:LM} and \eqref{eq:G} from the conformal Ward identity, and show that the same factor relates the replica partition function to $Z_L$ when a line is inserted on the $a$-cycle. After a conformal map, $A=[0,x]$ and $B=[1,\infty)$, and the replica surface is the curve $y^2=P(z)$ with $P(z)=z(z-x)(z-1)$. A flat coordinate on the torus is $u=\int dz/\sqrt{P(z)}$, and its periods along the $a$-cycle and the $b$-cycle are $\omega_a=4\mathbf K(x)$ and $\omega_b=4i\mathbf K(1-x)$, so that $\tau=\omega_b/\omega_a$.

The stress tensor of the orbifold is the sum of the stress tensors on the two sheets. Each of them is obtained from the stress tensor on the torus by the map $z\to u$, with $u'(z)^2=1/P(z)$, so that
\[
\frac{\langle T_{\rm orb}(z)\,\sigma(\infty)\sigma(1)\sigma(x)\sigma(0)\rangle}{G}=\frac{2\langle T(u)\rangle}{P(z)}+\frac c6\{u,z\},\qquad\{u,z\}=-\frac{P''}{2P}+\frac38\frac{P'^2}{P^2}.
\]
On the torus with periods $\omega_a$ and $\omega_b$, $\langle T(u)\rangle=(2\pi i/\omega_a^2)\,\partial_\tau\log Z$. Near $z=x$ the Schwarzian behaves as
\[
\{u,z\}=\frac{3/8}{(z-x)^2}-\frac{1}{4(z-x)}\,\frac{2x-1}{x(x-1)}+\dots .
\]
The double pole gives the weight $h_\sigma=\frac c6\cdot\frac38=\frac c{16}$ of the twist field. By the Ward identity the residue at $z=x$ is $\partial_x\log G$, and therefore
\[
\partial_x\log G=-\frac c{24}\,\frac{2x-1}{x(x-1)}+\frac{\pi i}{4\mathbf K(x)^2\,x(x-1)}\,\partial_\tau\log Z .
\]
The first term is $-\frac c{24}\partial_x\log(x(1-x))$. By the Legendre relation for elliptic integrals, $d\tau/dx=-i\pi/(4x(1-x)\mathbf K(x)^2)$, and the second term is $\partial_x\log Z$. Together with the corresponding equation for $\bar x$, this gives $G=\mathcal N|x(1-x)|^{-c/12}Z$. For $x\to0$ the operator product of $\sigma(x)$ and $\sigma(0)$ is dominated by the identity, so that $G\simeq|x|^{-c/4}$, while $Z\simeq|\fq|^{-c/12}\simeq|x/16|^{-c/6}$. Hence $\mathcal N=2^{-2c/3}$, which is \eqref{eq:G}. Writing the four-point function of the twist fields at the endpoints of $A$ and $B$ as $|z_{12}z_{34}|^{-c/4}$ times a function of $x$, and dividing by $\tr\rho_A^2\,\tr\rho_B^2$, gives \eqref{eq:LM}.

Now let a line $L$ run along the $a$-cycle. It commutes with the stress tensor and does not touch the branch points, so the argument goes through unchanged, except that the trace on the torus now contains $\hat L$ and $\langle T(u)\rangle=(2\pi i/\omega_a^2)\,\partial_\tau\log Z_L$. This gives $G_L=\mathcal N_L|x(1-x)|^{-c/12}Z_L$. For $x\to0$ the loop surrounds $\sigma(0)\sigma(x)$, whose operator product is dominated by the identity, around which the loop can be shrunk to give $d_L$. Since also $Z_L\simeq d_L|\fq|^{-c/12}$, the two factors of $d_L$ match and $\mathcal N_L=\mathcal N$. The ratio of the replica partition functions with and without the line is therefore $Z_L/Z$, as stated in \eqref{eq:obs}.

\section{Characters and identities}\label{app:data}

We collect here the functions used in section~\ref{sec:E8}. Let
\[
A(\fq)=\sum_{m,n\in\mathbb{Z}}\fq^{m^2+mn+n^2},\qquad
C(\fq)=\sum_{m,n\in\mathbb{Z}}\fq^{(m+\frac13)^2+(m+\frac13)(n+\frac13)+(n+\frac13)^2}=\frac{3\,\eta(3\tau)^3}{\eta(\tau)}
\]
be the theta functions of the $A_2$ lattice and of its translate by a fundamental weight. They satisfy $A^3=B^3+C^3$ with $B=\eta(\tau)^3/\eta(3\tau)$ \cite{Borwein:1991cubic}. Decomposing the $E_8$ lattice into cosets of $A_2\oplus E_6$, and the $E_6$ lattice into cosets of $A_2^{\oplus3}$, one finds
\[
\Theta_{E_6}=A^3+2C^3,\qquad\Theta_{E_6+\omega'}=3AC^2,\qquad\Theta_{E_8}=A^4+8AC^3 .
\]
The $\mathbb{Z}_3$ symmetry of section~\ref{sec:E8} acts as $e^{\pm2\pi i/3}$ on the two non-trivial cosets of $A_2\oplus E_6$ in $E_8$, which gives the twined character $(A^4-AC^3)/\eta^8$. Its defect Hilbert space is obtained from the shifted lattice $E_8+\omega$ \cite{Kac:1990gs} and has character $(A^3C+3A^2C^2+3AC^3+2C^4)/\eta^8$.

The characters of $(G_2)_1$ and $(F_4)_1$ are the solutions of the modular differential equation $\big(\vartheta^2-\frac{E_2}{6}\vartheta+\alpha_0\alpha_1E_4\big)\chi=0$, where $\vartheta=\fq\,\partial_\fq$, with exponents $(\alpha_0,\alpha_1)=(-\frac{7}{60},\frac{17}{60})$ for $G_2$ and $(-\frac{13}{60},\frac{23}{60})$ for $F_4$. With the leading coefficients normalized to $(1,7)$ and $(1,26)$ they are
\[
\chi^{G_2}_0=\fq^{-7/60}(1+14\fq+42\fq^2+\dots),\qquad\chi^{G_2}_{2/5}=\fq^{17/60}(7+34\fq+\dots),
\]
\[
\chi^{F_4}_0=\fq^{-13/60}(1+52\fq+377\fq^2+\dots),\qquad\chi^{F_4}_{3/5}=\fq^{23/60}(26+299\fq+\dots).
\]
We verified the relation $\chi_{E_8}=\chi_0\tilde\chi_0+\chi_{2/5}\tilde\chi_{3/5}$.

\section{Block coefficients}\label{app:blocks}

Tables~\ref{tab:s} and~\ref{tab:t} list the coefficients in \eqref{eq:master} for $(E_8)_1$ up to $H=14$, computed at $C=16$ with external weight $\frac12$. The first coefficients in the defect sectors are the multiplicities of the lowest levels of $\cH_L$, namely $16$ and $128$ for the $\mathbb{Z}_2$ line, $3$, $27$ and $81$ for the $\mathbb{Z}_3$ line, and $7$, $26$ and $182$ for $W$. At higher levels the Virasoro descendants of the orbifold theory have to be subtracted, and the coefficients are no longer integers. For instance $16^4a_4=\frac{203776}{51}$. Table~\ref{tab:towers} gives the combinations of blocks that correspond to a Virasoro module of the seed theory $T$ at $c=8$, as discussed in section~\ref{sec:blocks}.

\begin{table}[!htb]
\centering\small
\renewcommand{\arraystretch}{1.35}
\begin{tabular}{crrrr}
\toprule
$H$ & $16^H a_H$ & $16^H a^{g_2}_H$ & $16^H a^{g_3}_H$ & $16^H a^{W}_H$ \\
\midrule
$0$ & $1$ & $1$ & $1$ & $\varphi$ \\
$2$ & $248$ & $-8$ & $5$ & $182-116\varphi$ \\
$4$ & $\frac{203776}{51}$ & $\frac{50176}{1581}$ & $-\frac{15767}{1581}$ & $18.3010543$ \\
$6$ & $\frac{1150926848}{35505}$ & $-74.7707792$ & $3.74655183$ & $-66.6435052$ \\
$8$ & $192117.562$ & $152.279581$ & $16.3850098$ & $187.493387$ \\
$10$ & $922193.433$ & $-322.234334$ & $-41.3399483$ & $-369.40732$ \\
\bottomrule
\end{tabular}

\caption{The coefficients $16^Ha_H$ in the block expansion of the holomorphic twist-field correlator of $(E_8)_1$ at $C=16$, without a line and with each of the three lines on the $a$-cycle. The $\mathbb{Z}_2$ and $\mathbb{Z}_3$ lines are denoted by $g_2$ and $g_3$. Rational values are given when the denominators are small.}
\label{tab:s}
\vspace{0.8em}
\begin{tabular}{rrrrrr}
\toprule
\multicolumn{2}{c}{$\mathcal{H}_{g_2}$} & \multicolumn{2}{c}{$\mathcal{H}_{g_3}$} & \multicolumn{2}{c}{$\mathcal{H}_{W}$} \\
\cmidrule(lr){1-2}\cmidrule(lr){3-4}\cmidrule(lr){5-6}
$H$ & $16^H a^{(g_2)}_H$ & $H$ & $16^H a^{(g_3)}_H$ & $H$ & $16^H a^{(W)}_H$ \\
\midrule
$1$ & $16$ & $2/3$ & $3$ & $4/5$ & $7$ \\
$2$ & $128$ & $4/3$ & $27$ & $6/5$ & $26$ \\
$3$ & $\frac{5120}{9}$ & $2$ & $81$ & $2$ & $182$ \\
$4$ & $\frac{61440}{31}$ & $8/3$ & $\frac{20559}{85}$ & $14/5$ & $395.138686$ \\
$5$ & $\frac{15568896}{2599}$ & $10/3$ & $\frac{48222}{83}$ & $16/5$ & $650.794621$ \\
$6$ & $\frac{132448256}{8153}$ & $4$ & $\frac{41391}{31}$ & $4$ & $2883.06452$ \\
$7$ & $40718.1117$ & $14/3$ & $2786.23949$ & $24/5$ & $4311.93063$ \\
$8$ & $95982.6414$ & $16/3$ & $5636.89379$ & $26/5$ & $6614.21845$ \\
\bottomrule
\end{tabular}

\caption{The coefficients for the three lines on the $b$-cycle, for which the exchanged states belong to $\cH_L$ and $H=2h$.}
\label{tab:t}
\end{table}

\begin{table}[!htb]
\centering\footnotesize
\setlength{\tabcolsep}{4.5pt}\renewcommand{\arraystretch}{1.2}
\begin{tabular}{lccccccc}
\toprule
seed module & $k=0$ & $k=1$ & $k=2$ & $k=3$ & $k=4$ & $k=5$ & $k=6$\\
\midrule
vacuum & 1.0000 & $\mathbf{0}$ & 0.6078 & 0.2296 & 0.6177 & 0.1710 & 1.0095\\
$h\to0^+$ & 1.0000 & 1.0000 & 1.0917 & 1.1463 & 1.7317 & 1.8319 & 2.8274\\
$h=\frac14$ & 1.0000 & 0.6774 & 1.0281 & 1.1427 & 1.7111 & 1.8292 & 2.8066\\
$h=1$ & 1.0000 & 0.4839 & 0.9167 & 1.1140 & 1.6609 & 1.8179 & 2.7562\\
$h=3$ & 1.0000 & 0.4696 & 0.8374 & 1.0523 & 1.5839 & 1.7833 & 2.6732\\
\bottomrule
\end{tabular}

\caption{The coefficients $16^Hb_k$ with which a Virasoro module of weight $h$ of a theory with $c=8$ contributes to the blocks with $H=2h+2k$. For the vacuum module the coefficient of the block with $H=2$ vanishes.}
\label{tab:towers}
\end{table}

\section{Details of the lattice computation}\label{app:lattice}

We use the Majorana operators $a_{2j-1}=\big(\prod_{k<j}\sigma^z_k\big)\sigma^x_j$ and $a_{2j}=\big(\prod_{k<j}\sigma^z_k\big)\sigma^y_j$, in terms of which $\sigma^z_j=-ia_{2j-1}a_{2j}$ and $\sigma^x_j\sigma^x_{j+1}=-ia_{2j}a_{2j+1}$. A Gaussian operator $\rho_\Gamma$ is specified by a real antisymmetric matrix $\Gamma$ through $\langle a_ma_n\rangle=\delta_{mn}+i\Gamma_{mn}$ and Wick's theorem. In the ground state of the critical chain the non-vanishing elements are $\Gamma_{2j-1,2l}=-\Gamma_{2l,2j-1}=2/(\pi(2(l-j)+1))$. For two Gaussian operators on $N$ sites,
\[
\tr\big(\rho_{\Gamma_1}\rho_{\Gamma_2}\big)=(-1)^N2^{-N}\operatorname{Pf}\begin{pmatrix}\Gamma_1&\mathbbm 1\\-\mathbbm 1&-\Gamma_2\end{pmatrix},
\]
and conjugation by the fermion parity $P_X$ of a set of sites $X$ changes the sign of the rows and columns of $\Gamma$ that belong to $X$. Let $\Gamma_{AB}$ be the restriction of $\Gamma$ to the two blocks, let $G$ denote the sites between them, and let $\Gamma'=\Gamma_{AB}-\Gamma_{AB,G}\Gamma_{GG}^{-1}\Gamma_{G,AB}$ and $w=\operatorname{Pf}\Gamma_{GG}$, which is the expectation value of $P_G$. Then \cite{Fagotti:2010yr}
\[
\rho_{AB}=\tfrac12\big(\rho_{\Gamma_{AB}}+P_B\,\rho_{\Gamma_{AB}}P_B\big)+\tfrac w2\big(\rho_{\Gamma'}-P_B\,\rho_{\Gamma'}P_B\big),
\]
and each of the two traces in the lattice overlap is a sum of sixteen terms of the form given above. Table~\ref{tab:fits} shows how the extrapolated values of $\cO_\eta$ depend on the range of $\ell$ and on the terms included in the fit.

\begin{table}[!htb]
\centering\small
\renewcommand{\arraystretch}{1.2}
\begin{tabular}{lcccc}
\toprule
fit & $x=\frac19$ & $x=\frac14$ & $x=\frac49$ & $x=\frac{16}{25}$\\
\midrule
$16\le\ell\le128$, terms $\ell^{-1/2},\ell^{-1}$ & $0.54681$ & $0.46376$ & $0.39052$ & $0.32955$\\
$32\le\ell\le128$, terms $\ell^{-1/2},\ell^{-1}$ & $0.54685$ & $0.46381$ & $0.39058$ & $0.32966$\\
$16\le\ell\le96$, terms $\ell^{-1/2},\ell^{-1}$ & $0.54680$ & $0.46375$ & $0.39049$ & $0.32950$\\
$8\le\ell\le128$, terms $\ell^{-1/2},\ell^{-1},\ell^{-3/2}$ & $0.54689$ & $0.46386$ & $0.39067$ & $0.32984$\\
$16\le\ell\le128$, terms $\ell^{-1/2},\ell^{-1},\ell^{-3/2}$ & $0.54688$ & $0.46385$ & $0.39065$ & $0.32980$\\
\midrule
equation \eqref{eq:isingI2} & $0.54688$ & $0.46385$ & $0.39064$ & $0.32977$\\
\bottomrule
\end{tabular}

\caption{Extrapolated lattice values of $\cO_\eta$ for different fits of the data to a constant plus the indicated powers of $\ell$. The first line is the fit used in table~\ref{tab:ising}.}
\label{tab:fits}
\end{table}

\clearpage
\bibliographystyle{JHEP}
\bibliography{refs}

\providecommand{\href}[2]{#2}\begingroup\raggedright\begin{thebibliography}{10}

\bibitem{Calabrese:2004eu}
P.~Calabrese and J.~Cardy, {\it {Entanglement entropy and quantum field
  theory}},  {\em J. Stat. Mech.} {\bf 0406} (2004) P06002,
  [\href{http://arxiv.org/abs/hep-th/0405152}{{\tt hep-th/0405152}}].

\bibitem{Furukawa:2008uk}
S.~Furukawa, V.~Pasquier, and J.~Shiraishi, {\it {Mutual Information and
  Compactification Radius in a c=1 Critical Phase in One Dimension}},  {\em
  Phys. Rev. Lett.} {\bf 102} (2009) 170602,
  [\href{http://arxiv.org/abs/0809.5113}{{\tt arXiv:0809.5113}}].

\bibitem{Calabrese:2009ez}
P.~Calabrese, J.~Cardy, and E.~Tonni, {\it {Entanglement entropy of two
  disjoint intervals in conformal field theory}},  {\em J. Stat. Mech.} {\bf
  0911} (2009) P11001, [\href{http://arxiv.org/abs/0905.2069}{{\tt
  arXiv:0905.2069}}].

\bibitem{Calabrese:2010he}
P.~Calabrese, J.~Cardy, and E.~Tonni, {\it {Entanglement entropy of two
  disjoint intervals in conformal field theory II}},  {\em J. Stat. Mech.} {\bf
  1101} (2011) P01021, [\href{http://arxiv.org/abs/1011.5482}{{\tt
  arXiv:1011.5482}}].

\bibitem{Headrick:2010zt}
M.~Headrick, {\it {Entanglement R\'enyi entropies in holographic theories}},
  {\em Phys. Rev. D} {\bf 82} (2010) 126010,
  [\href{http://arxiv.org/abs/1006.0047}{{\tt arXiv:1006.0047}}].

\bibitem{Lunin:2000yv}
O.~Lunin and S.~D. Mathur, {\it {Correlation functions for $M^N/S_N$
  orbifolds}},  {\em Commun. Math. Phys.} {\bf 219} (2001) 399--442,
  [\href{http://arxiv.org/abs/hep-th/0006196}{{\tt hep-th/0006196}}].

\bibitem{Frohlich:2006ch}
J.~Fr\"ohlich, J.~Fuchs, I.~Runkel, and C.~Schweigert, {\it {Duality and
  defects in rational conformal field theory}},  {\em Nucl. Phys. B} {\bf 763}
  (2007) 354--430, [\href{http://arxiv.org/abs/hep-th/0607247}{{\tt
  hep-th/0607247}}].

\bibitem{Bhardwaj:2017xup}
L.~Bhardwaj and Y.~Tachikawa, {\it {On finite symmetries and their gauging in
  two dimensions}},  {\em JHEP} {\bf 03} (2018) 189,
  [\href{http://arxiv.org/abs/1704.02330}{{\tt arXiv:1704.02330}}].

\bibitem{Chang:2018iay}
C.-M. Chang, Y.-H. Lin, S.-H. Shao, Y.~Wang, and X.~Yin, {\it {Topological
  Defect Lines and Renormalization Group Flows in Two Dimensions}},  {\em JHEP}
  {\bf 01} (2019) 026, [\href{http://arxiv.org/abs/1802.04445}{{\tt
  arXiv:1802.04445}}].

\bibitem{Casini:2019kex}
H.~Casini, M.~Huerta, J.~M. Magan, and D.~Pontello, {\it {Entanglement entropy
  and superselection sectors. Part I. Global symmetries}},  {\em JHEP} {\bf 02}
  (2020) 014, [\href{http://arxiv.org/abs/1905.10487}{{\tt arXiv:1905.10487}}].

\bibitem{Benedetti:2024dku}
V.~Benedetti, H.~Casini, Y.~Kawahigashi, R.~Longo, and J.~M. Magan, {\it
  {Modular invariance as completeness}},
  \href{http://arxiv.org/abs/2408.04011}{{\tt arXiv:2408.04011}}.

\bibitem{Benedetti:2026jones}
V.~Benedetti, I.~Davila-Cuba, and E.~Tonni, {\it {Jones index from R\'enyi
  entropies in the Ising conformal field theory}},  {\em JHEP} {\bf 06} (2026)
  241, [\href{http://arxiv.org/abs/2603.13013}{{\tt arXiv:2603.13013}}].

\bibitem{Ares:2022gjb}
F.~Ares, P.~Calabrese, G.~Di~Giulio, and S.~Murciano, {\it {Multi-charged
  moments of two intervals in conformal field theory}},  {\em JHEP} {\bf 09}
  (2022) 051, [\href{http://arxiv.org/abs/2206.01534}{{\tt arXiv:2206.01534}}].

\bibitem{Bruno:2023tez}
A.~Bruno, F.~Ares, S.~Murciano, and P.~Calabrese, {\it {Symmetry resolution of
  the computable cross-norm negativity of two disjoint intervals in the
  massless Dirac field theory}},  \href{http://arxiv.org/abs/2312.02926}{{\tt
  arXiv:2312.02926}}.

\bibitem{Coser:2015dvp}
A.~Coser, E.~Tonni, and P.~Calabrese, {\it {Spin structures and entanglement of
  two disjoint intervals in conformal field theories}},  {\em J. Stat. Mech.}
  {\bf 1605} (2016) 053109, [\href{http://arxiv.org/abs/1511.08328}{{\tt
  arXiv:1511.08328}}].

\bibitem{Goldstein:2017bua}
M.~Goldstein and E.~Sela, {\it {Symmetry-resolved entanglement in many-body
  systems}},  {\em Phys. Rev. Lett.} {\bf 120} (2018) 200602,
  [\href{http://arxiv.org/abs/1711.09418}{{\tt arXiv:1711.09418}}].

\bibitem{Xavier:2018kqb}
J.~C. Xavier, F.~C. Alcaraz, and G.~Sierra, {\it {Equipartition of the
  entanglement entropy}},  {\em Phys. Rev. B} {\bf 98} (2018) 041106,
  [\href{http://arxiv.org/abs/1804.06357}{{\tt arXiv:1804.06357}}].

\bibitem{Saura-Bastida:2024yye}
P.~Saura-Bastida, A.~Das, G.~Sierra, and J.~Molina-Vilaplana, {\it
  {Categorical-symmetry resolved entanglement in CFT}},  {\em Phys. Rev. D}
  {\bf 109} (2024) 105026, [\href{http://arxiv.org/abs/2402.06322}{{\tt
  arXiv:2402.06322}}].

\bibitem{Das:2024qdx}
A.~Das, J.~Molina-Vilaplana, and P.~Saura-Bastida, {\it {Generalized symmetry
  resolution of entanglement in CFT for twisted and anyonic sectors}},  {\em
  Phys. Rev. D} {\bf 110} (2024) 125005,
  [\href{http://arxiv.org/abs/2409.02162}{{\tt arXiv:2409.02162}}].

\bibitem{Ares:2022koq}
F.~Ares, S.~Murciano, and P.~Calabrese, {\it {Entanglement asymmetry as a probe
  of symmetry breaking}},  {\em Nature Commun.} {\bf 14} (2023) 2036,
  [\href{http://arxiv.org/abs/2207.14693}{{\tt arXiv:2207.14693}}].

\bibitem{Benini:2025asym}
F.~Benini, P.~Calabrese, M.~Fossati, A.~H. Singh, and M.~Venuti, {\it
  {Entanglement asymmetry for higher and noninvertible symmetries}},
  \href{http://arxiv.org/abs/2509.16311}{{\tt arXiv:2509.16311}}.

\bibitem{Lin:2019kpn}
Y.-H. Lin and S.-H. Shao, {\it {Anomalies and Bounds on Charged Operators}},
  {\em Phys. Rev. D} {\bf 100} (2019) 025013,
  [\href{http://arxiv.org/abs/1904.04833}{{\tt arXiv:1904.04833}}].

\bibitem{Umasankar:2026tdl}
N.~B. Umasankar and A.~Das, {\it {Verlinde lines, anyon permutations and
  commutant pairs inside $E_{8,1}$ CFT}}, . to appear.

\bibitem{Albert:2025haag}
J.~Albert, Y.~Honda, J.~Kaidi, and Y.~Zheng, {\it {Haagerup Symmetry in
  $(E_8)_1$?}},  {\em Phys. Rev. Lett.} {\bf 136} (2026) 091603,
  [\href{http://arxiv.org/abs/2512.08225}{{\tt arXiv:2512.08225}}].

\bibitem{Ruggiero:2018hyl}
P.~Ruggiero, E.~Tonni, and P.~Calabrese, {\it {Entanglement entropy of two
  disjoint intervals and the recursion formula for conformal blocks}},  {\em J.
  Stat. Mech.} {\bf 1811} (2018) 113101,
  [\href{http://arxiv.org/abs/1805.05975}{{\tt arXiv:1805.05975}}].

\bibitem{Poland:2026vb}
D.~Poland, {\it {virasoro\_blocks: Software for the 2d bootstrap}},
  \href{http://arxiv.org/abs/2610.39106}{{\tt arXiv:2610.39106}}.

\bibitem{Lin:2021udi}
Y.-H. Lin and S.-H. Shao, {\it {$\mathbb{Z}_N$ symmetries, anomalies, and the
  modular bootstrap}},  {\em Phys. Rev. D} {\bf 103} (2021) 125001,
  [\href{http://arxiv.org/abs/2101.08343}{{\tt arXiv:2101.08343}}].

\bibitem{Seiberg:2023cdc}
N.~Seiberg and S.-H. Shao, {\it {Majorana chain and Ising model --
  (non-invertible) translations, anomalies, and emanant symmetries}},  {\em
  SciPost Phys.} {\bf 16} (2024) 064,
  [\href{http://arxiv.org/abs/2307.02534}{{\tt arXiv:2307.02534}}].

\bibitem{Lin:2022dhv}
Y.-H. Lin, M.~Okada, S.~Seifnashri, and Y.~Tachikawa, {\it {Asymptotic density
  of states in 2d CFTs with non-invertible symmetries}},  {\em JHEP} {\bf 03}
  (2023) 094, [\href{http://arxiv.org/abs/2208.05982}{{\tt arXiv:2208.05982}}].

\bibitem{AliAhmad:2025faces}
S.~Ali~Ahmad, M.~S. Klinger, and Y.~Wang, {\it {The Many Faces of
  Non-invertible Symmetries}},  \href{http://arxiv.org/abs/2509.18072}{{\tt
  arXiv:2509.18072}}.

\bibitem{Petz:1988suff}
D.~Petz, {\it {Sufficiency of channels over von Neumann algebras}},  {\em
  Quart. J. Math. Oxford} {\bf 39} (1988) 97--108.

\bibitem{Borwein:1991cubic}
J.~M. Borwein and P.~B. Borwein, {\it {A cubic counterpart of Jacobi's identity
  and the AGM}},  {\em Trans. Amer. Math. Soc.} {\bf 323} (1991) 691--701.

\bibitem{Mathur:1988na}
S.~D. Mathur, S.~Mukhi, and A.~Sen, {\it {On the Classification of Rational
  Conformal Field Theories}},  {\em Phys. Lett. B} {\bf 213} (1988) 303.

\bibitem{Naculich:1988xv}
S.~G. Naculich, {\it {Differential equations for rational conformal
  characters}},  {\em Nucl. Phys. B} {\bf 323} (1989) 423--440.

\bibitem{Alba:2009ek}
V.~Alba, L.~Tagliacozzo, and P.~Calabrese, {\it {Entanglement entropy of two
  disjoint blocks in critical Ising models}},  {\em Phys. Rev. B} {\bf 81}
  (2010) 060411, [\href{http://arxiv.org/abs/0910.0706}{{\tt
  arXiv:0910.0706}}].

\bibitem{Alba:2011fu}
V.~Alba, L.~Tagliacozzo, and P.~Calabrese, {\it {Entanglement entropy of two
  disjoint intervals in c=1 theories}},  {\em J. Stat. Mech.} {\bf 1106} (2011)
  P06012, [\href{http://arxiv.org/abs/1103.3166}{{\tt arXiv:1103.3166}}].

\bibitem{Fagotti:2010yr}
M.~Fagotti and P.~Calabrese, {\it {Entanglement entropy of two disjoint blocks
  in XY chains}},  {\em J. Stat. Mech.} {\bf 1004} (2010) P04016,
  [\href{http://arxiv.org/abs/1003.1110}{{\tt arXiv:1003.1110}}].

\bibitem{Cardy:2010zs}
J.~Cardy and P.~Calabrese, {\it {Unusual corrections to scaling in entanglement
  entropy}},  {\em J. Stat. Mech.} {\bf 1004} (2010) P04023,
  [\href{http://arxiv.org/abs/1002.4353}{{\tt arXiv:1002.4353}}].

\bibitem{Tambara:1998vmj}
D.~Tambara and S.~Yamagami, {\it {Tensor categories with fusion rules of
  self-duality for finite abelian groups}},  {\em J. Algebra} {\bf 209} (1998)
  692--707.

\bibitem{Kitaev:2005hzj}
A.~Kitaev, {\it {Anyons in an exactly solved model and beyond}},  {\em Annals
  Phys.} {\bf 321} (2006) 2--111,
  [\href{http://arxiv.org/abs/cond-mat/0506438}{{\tt cond-mat/0506438}}].

\bibitem{Witten:1988hf}
E.~Witten, {\it {Quantum field theory and the Jones polynomial}},  {\em Commun.
  Math. Phys.} {\bf 121} (1989) 351--399.

\bibitem{Fuchs:2002cm}
J.~Fuchs, I.~Runkel, and C.~Schweigert, {\it {TFT construction of RCFT
  correlators I: Partition functions}},  {\em Nucl. Phys. B} {\bf 646} (2002)
  353--497, [\href{http://arxiv.org/abs/hep-th/0204148}{{\tt hep-th/0204148}}].

\bibitem{Kalmeyer:1987vb}
V.~Kalmeyer and R.~B. Laughlin, {\it {Equivalence of the
  resonating-valence-bond and fractional quantum Hall states}},  {\em Phys.
  Rev. Lett.} {\bf 59} (1987) 2095--2098.

\bibitem{Umasankar:2026pv}
N.~B. Umasankar, M.~S. Mitchell, and D.~Poland, ``{3d CFT Bootstrap with Parity
  Violation}.'' to appear, 2026.

\bibitem{Mignard2017ModularCA}
M.~Mignard and P.~Schauenburg, {\it {Modular categories are not determined by
  their modular data}},  {\em Lett. Math. Phys.} {\bf 111} (2021) 60,
  [\href{http://arxiv.org/abs/1708.02796}{{\tt arXiv:1708.02796}}].

\bibitem{Zamolodchikov:1984eqp}
A.~B. Zamolodchikov, {\it {Conformal symmetry in two dimensions: an explicit
  recurrence formula for the conformal partial wave amplitude}},  {\em Commun.
  Math. Phys.} {\bf 96} (1984) 419--422.

\bibitem{Zamolodchikov:1987avt}
A.~B. Zamolodchikov, {\it {Conformal symmetry in two-dimensional space:
  recursion representation of conformal block}},  {\em Theor. Math. Phys.} {\bf
  73} (1987) 1088--1093.

\bibitem{Headrick:2015gba}
M.~Headrick, A.~Maloney, E.~Perlmutter, and I.~G. Zadeh, {\it {R\'enyi
  entropies, the analytic bootstrap, and 3D quantum gravity at higher genus}},
  \href{http://arxiv.org/abs/1503.07111}{{\tt arXiv:1503.07111}}.

\bibitem{Birman1973Isotopies}
J.~S. Birman and H.~M. Hilden, {\it {On isotopies of homeomorphisms of Riemann
  surfaces}},  {\em Ann. Math.} {\bf 97} (1973) 424--439.

\bibitem{Wen:2019ylt}
X.~Wen and X.-G. Wen, {\it {Distinguish modular categories and 2+1D topological
  orders beyond modular data: mapping class group of higher genus manifold}},
  \href{http://arxiv.org/abs/1908.10381}{{\tt arXiv:1908.10381}}.

\bibitem{Huang:2021nvb}
T.-C. Huang, Y.-H. Lin, K.~Ohmori, Y.~Tachikawa, and M.~Tezuka, {\it {Numerical
  Evidence for a Haagerup Conformal Field Theory}},  {\em Phys. Rev. Lett.}
  {\bf 128} (2022) 231603, [\href{http://arxiv.org/abs/2110.03008}{{\tt
  arXiv:2110.03008}}].

\bibitem{Vanhove:2021zop}
R.~Vanhove, L.~Lootens, M.~Van~Damme, R.~Wolf, T.~J. Osborne, J.~Haegeman, and
  F.~Verstraete, {\it {Critical Lattice Model for a Haagerup Conformal Field
  Theory}},  {\em Phys. Rev. Lett.} {\bf 128} (2022) 231602,
  [\href{http://arxiv.org/abs/2110.03532}{{\tt arXiv:2110.03532}}].

\bibitem{Evans:2010yr}
D.~E. Evans and T.~Gannon, {\it {The exoticness and realisability of twisted
  Haagerup--Izumi modular data}},  {\em Commun. Math. Phys.} {\bf 307} (2011)
  463--512, [\href{http://arxiv.org/abs/1006.1326}{{\tt arXiv:1006.1326}}].

\bibitem{Lin:2019hks}
Y.-H. Lin and S.-H. Shao, {\it {Duality Defect of the Monster CFT}},  {\em J.
  Phys. A} {\bf 54} (2021) 065201, [\href{http://arxiv.org/abs/1911.00042}{{\tt
  arXiv:1911.00042}}].

\bibitem{Kac:1990gs}
V.~G. Kac, {\em {Infinite dimensional Lie algebras}}.
\newblock Cambridge University Press, 3~ed., 1990.

\end{thebibliography}\endgroup
\end{document}